\documentclass{article}

\usepackage{PRIMEarxiv}

\usepackage[utf8]{inputenc} 
\usepackage[T1]{fontenc}    
\usepackage{hyperref}       
\usepackage{url}            
\usepackage{booktabs}       
\usepackage{amsfonts}       
\usepackage{nicefrac}       
\usepackage{microtype}      
\usepackage{lipsum}
\usepackage{fancyhdr}       
\usepackage{graphicx}       
\usepackage{amsmath,amssymb}
\usepackage{xcolor}
\usepackage{subcaption}
\usepackage{siunitx} 
\usepackage{longtable}
\usepackage{authblk}
\newcommand{\hide}[1]{}
\newcommand{\etal}{\textit{\emph{et al.}}.}

\title{Image-Based Whole-Heart Cardiac Flow Simulations in Health and Congenital Heart Disease
}
\author{
  Fanwei Kong$^{1,2,3,4}$ \quad
  Aaron Brown$^{5}$ \quad
  Michael Loecher$^{6}$ \quad
  Perry S. Choi$^{7}$ \quad
  Lei Shi$^{8}$ \quad
  Michael Ma$^{7}$ \quad
  Daniel B. Ennis$^{1,6,9}$ \quad
  Alison Marsden$^{1,4,5,9}$ \\
  {\small
  $^{1}$Cardiovascular Institute, Stanford University \\
  $^{2}$Mechanical Engineering \& Materials Science, Washington University in St. Louis \\
  $^{3}$Department of Pediatrics, Stanford University \\
  $^{4}$Institute for Computational and Mathematical Engineering, Stanford University \\
  $^{5}$Department of Mechanical Engineering, Stanford University \\
  $^{6}$Department of Radiology, Stanford University \\
  $^{7}$Department of Cardiothoracic Surgery, Stanford University \\
  $^{8}$Department of Mechanical Engineering, Kennesaw State University \\
  $^{9}$Department of Bioengineering, Stanford University \\
  }
}

\begin{document}
\maketitle

\begin{abstract}
Intracardiac flow patterns are shaped by the coupled motion of the cardiac chambers and heart valves and provide important information about cardiac function. However, clinical flow imaging remains limited by exam times, noise, resolution, and incomplete details of the three-dimensional flow. Computational fluid dynamics (CFD) can potentially provide detailed flow quantification and predictive insight into treatment outcomes, but clinical translation requires frameworks that reproduce patient-specific measurements while balancing physiological realism, computational cost, and modeling effort. Herein, we present an image-based, patient-specific computational framework for simulating whole-heart intracardiac hemodynamics that balances physiological fidelity with computational efficiency. The framework first employs machine learning-based segmentation and mesh propagation to reconstruct moving cardiac anatomies from time-resolved images. CFD simulations are then performed to resolve blood flow in deforming domains, while resistive immersed surfaces (RIS) are used to model all four cardiac valves with physiologically realistic opening and closing dynamics. The framework was applied to model hemodynamics in a healthy adult and a pediatric patient with complex congenital heart disease (CHD). In the healthy case, the simulations reproduced physiologic pressure-volume behavior, valve timing, and ventricular vortex formation. In the CHD case, simulated chamber and vessel pressures showed agreement with cardiac catheterization measurements. Simulated flow fields were qualitatively consistent with 4D-Flow MRI, while providing higher-resolution visualization of flow structures that were partially obscured by imaging artifacts. Comparison between the healthy and CHD cases further revealed altered diastolic flow organization and elevated normalized viscous dissipation in the CHD heart. 

\end{abstract}


\section{Introduction}

Intracardiac flow dynamics are driven by the coupled motion between the heart valves and cardiac chambers. These coordinated motions generate complex, time-varying flow structures that facilitate efficient blood transport and energy exchange within the heart. For example, during the filling phase of a cardiac cycle, vortices arise as blood transitions from narrow inflow tracts into the larger ventricular cavities\cite{kilner_asymmetric_2000,mangual_describing_2012,rodriguez_munoz_intracardiac_2013}, and play a critical role in minimizing energy dissipation, preserving momentum, and directing blood efficiently toward the outflow tract\cite{kilner_asymmetric_2000,mangual_describing_2012,rodriguez_munoz_intracardiac_2013,pedrizzetti_nature_2005,seo_effect_2013,cimino_vivo_2012}. In disease states, abnormal cardiac motion or valvular dysfunction can alter flow structures, leading to inefficient transport and impaired cardiac performance. Indeed, quantitative assessments of flow topology, vortex dynamics, and energy dissipation have been linked to the pathogenesis of various cardiovascular diseases, such as heart failure \cite{davies_abnormal_1992,ohno_mechanism_1994,cicchitti_heart_2016,carlhall_passing_2010}, acute myocardial infarction\cite{mcgarvey_directed_2013,chan_impact_2019}, valvular heart disease\cite{pedrizzetti_left_2010}, and dilated or hypertrophic cardiomyopathy\cite{gharib_optimal_2006,pedrizzetti_vortexearly_2014}. Thus, characterizing these ventricular flow patterns offers unique insights into patient cardiac function, the initiation and progression of cardiac diseases, and the development of treatment strategies that improve patient hemodynamics.

A variety of imaging modalities have been used to quantify intracardiac blood flow, including Doppler echocardiography, echocardiographic particle image velocimetry, and emerging ultrasound-based approaches such as blood speckle imaging \cite{russ_pulsed_2023, prinz_can_2012, assi_intraventricular_2017, nyrnes_blood_2020}. More recently, phase-contrast MRI and 4D-Flow MRI have gained popularity for providing time-resolved velocity fields that capture cardiac flow dynamics throughout the cardiac cycle, although they require long acquisition times and patient compliance \cite{bissell_4d_2023, azarine_four-dimensional_2019, demirkiran_clinical_2022}. Despite these advances, clinical flow imaging remains limited by spatial and temporal resolution, noise, and reconstruction artifacts \cite{russ_pulsed_2023, assi_intraventricular_2017, nyrnes_blood_2020, bissell_4d_2023, demirkiran_clinical_2022}, restricting its ability to accurately resolve detailed flow features such as vortices and wall shear stress that are critical for understanding disease mechanisms and guiding therapeutic interventions \cite{mazzi_revised_2025,dong_analysis_2022,schmidt_impact_2021,comunale_ventricular_2022,maire_abnormalities_1994,elbaz_impact_2015}.

Computational fluid dynamics (CFD) simulations have been widely used to investigate detailed intraventricular flow patterns arising from the dynamic motion of the beating heart. The fidelity of these simulations strongly depends on accurate representation of cardiac wall motion and heart valve dynamics, which govern inflow, outflow, and downstream flow structures, as well as physiologically appropriate inlet and outlet boundary conditions that model the surrounding circulatory system. Cardiac motion in CFD models is typically prescribed using either image-based or physics-based approaches. In image-based methods, wall motion is extracted directly from time-resolved imaging such as cine MRI or echocardiography and imposed as moving boundary conditions \cite{kong_learning_2023, kong_automating_2020, seo_effect_2013, pedrizzetti_left_2010, vedula_hemodynamics_2015, chnafa_image-based_2014, mao_fully-coupled_2017}. In comparison, physics-based approaches derive cardiac motion from electromechanical models that solve electrophysiology and myocardial mechanics as part of a fully coupled fluid–structure interaction (FSI) framework, allowing bidirectional coupling between cardiac deformation and blood flow \cite{quarteroni_integrated_2017, davey_simulating_2024, feng_whole-heart_2024, kariya_personalized_2020, viola_high-fidelity_2023, zingaro_electromechanics-driven_2024, bucelli_mathematical_2023}. Heart valves play a central role in intraventricular flow organization, particularly during diastole when atrioventricular valve dynamics govern inflow direction and vortex formation. While some studies have incorporated valves within fully coupled FSI models \cite{mao_fully-coupled_2017, davey_simulating_2024, feng_whole-heart_2024, kaiser_comparison_2023, kaiser_fluid-structure_2025}, these approaches remain computationally expensive. Alternatively, resistive immersed surface (RIS) methods provide a more efficient representation of valve dynamics through penalty-based resistance terms \cite{zingaro_modeling_2023, this_augmented_2020, fedele_patient-specific_2017}.

Recent studies have modeled whole-heart cardiac flow using frameworks with varying levels of physiological and numerical fidelity. However, these approaches have been applied either to idealized average cardiac anatomies or to single-patient anatomies reconstructed from imaging data, and without full personalization in which both wall motion and inlet/outlet boundary conditions are calibrated against clinical data and compared against flow imaging data. For example, Davey \etal \cite{davey_simulating_2024} and Feng \etal \cite{feng_whole-heart_2024} developed fully coupled fluid–structure interaction models within the immersed boundary framework \cite{peskin_immersed_2002}, incorporating biomechanically detailed representations of all major cardiac structures, including all four valves. In contrast, Zingaro \etal \cite{zingaro_electromechanics-driven_2024} proposed a more computationally efficient approach that nonetheless captures physiologically realistic cardiac flow. Their framework employs an Arbitrary Lagrangian–Eulerian (ALE) formulation \cite{noauthor_arbitrary_nodate,hirt_arbitrary_1974} with valves represented as resistive immersed surfaces in an average healthy heart geometry, where cardiac motion is prescribed from a separate electromechanical model simulating contraction and relaxation under both healthy and pathological conditions. A similar strategy of using ALE-based cardiac flow simulations with motion prescribed from mechanics models has also been adopted by the UT-Heart project \cite{okada_clinical_2019}. In addition, Karabelas \etal \cite{karabelas_global_2022} constructed whole-heart anatomies and imposed cardiac motion derived directly from dynamic patient imaging. They simulated cardiac flow using the ALE framework, representing all four heart valves, which were not visible in the images, as simplified planar structures, and modeling valve closure by introducing a Darcy-type drag term in the Navier–Stokes equations to penalize flow within the valve regions. 

Motivated by the goal of using CFD to obtain detailed, patient-specific cardiac flow information, we seek a modeling framework that balances simulation fidelity, computational cost, and human effort. Such a framework should yield simulation results on clinically relevant timescales, reproduce available patient measurements (e.g., pressures and flow imaging), and provide additional spatiotemporally resolved flow information that is not directly accessible clinically. While fully coupled FSI models offer the potential to simulate cardiac contraction and function as patient-specific digital twins, they remain computationally expensive and require substantial parameter tuning to achieve agreement with clinical data. Therefore, in this study, we adopt an image--based approach in which cardiac motion is efficiently derived from time-resolved imaging using machine learning--based segmentation and diffeomorphic registration driven by neural ordinary differential equations \cite{chen_neural_2018}, and prescribed as boundary conditions for the CFD solver. The resulting moving-domain flow problem is solved using the ALE formulation \cite{noauthor_arbitrary_nodate,hirt_arbitrary_1974} to accommodate large cardiac deformations over the cardiac cycle. We further extend the RIS method to model all four cardiac valves within patient-specific whole-heart anatomies using realistic valve geometries, with valve motion prescribed from pressure-driven mechanical simulations. To represent the interaction with the systemic and pulmonary circulations, the 3D CFD model is coupled to a closed-loop lumped parameter network. Together, this framework enables computationally efficient, patient-specific simulation of intracardiac flow. The proposed approach is implemented within the SimVascular/svFSI project \cite{zhu_svfsi_2022}, an open-source finite element platform for multiphysics cardiovascular simulations.

Congenital heart defects encompass a wide range of anatomical abnormalities present at birth, many of which can significantly disrupt normal cardiac flow dynamics \cite{marchese_left_2021, vasanawala_congenital_2015}. Malformations such as complex ventricular septal defects and single-ventricle physiology can substantially alter pressure and velocity fields, leading to inefficient circulation and an increased risk of morbidity and mortality. Such lesions often require complex surgeries in which an intracardiac baffle patch is inserted to restore physiologic bi-ventricular circulation \cite{liang_feasibility_2022},  \cite{nakamura_patient-specific_2021, mendez_apical_2018},  \cite{chen_digital_2018}. However, these complex surgeries remain non-standardized, still dependent on the surgeon's experience and intraoperative judgment, and commonly require reoperations or reinterventions \cite{sun_cardiovascular_2024}. Understanding the intricate flow patterns in such pathologies could enhance preoperative surgical planning and minimize the guesswork required. Intracardiac hemodynamics are an important consideration when designing individualized surgical plans aimed at optimizing cardiac function and hemodynamic outcomes.  While some CFD studies have been used to support surgical planning in CHD, they have primarily focused on the great arteries (i.e., the aorta or pulmonary arteries) \cite{kaiser_simulation-based_2024, pennati_computational_2013,marsden_computational_2015,vignon-clementel_primer_2010,marsden_evaluation_2009, capelli_patient-specific_2017, qian_computational_2010,dubini_numerical_1996, jack_numerical_2024}, with limited attention given to simulating intracardiac flow within malformed ventricular chambers. In the cardiac domain, Tang \emph{et al.}.\cite{tang_image-based_2010} applied FSI modeling to simulate cardiac fluid dynamics in a patient-specific right ventricle (RV) of a pediatric patient with Tetralogy of Fallot (ToF), evaluating the resulting stress and strain patterns under two different ventricular patch designs. Loke \emph{et al.}. \cite{loke_computational_2022} employed an immersed-boundary-based CFD solver to simulate the RV flow patterns in both normal controls and in patients with ToF, aiming to identify flow biomarkers that could potentially guide the timing of pulmonary valve replacement in ToF patients.  However, both studies constructed models from cine MRI data with limited spatial resolution and did not incorporate the critical influence of heart valves on intraventricular flow patterns. Their analysis was also limited to RV flow and did not include flow within the full four-chamber heart. The UT-Heart project \cite{kariya_personalized_2020, sugiura_ut-heart_2022} developed a multiphysics finite element solver for coupled electromechanical and FSI simulations in a complex CHD patient with double-outlet right ventricle (DORV) accompanied by ventricular and atrial septal defects. Their framework also evaluated the effects of surgical correction on intracardiac pressure and flow distributions. However, the study did not investigate detailed intraventricular flow pattern abnormalities or validate flow structures against 4D-Flow MRI data. 

Here, we simulate detailed whole-heart cardiac blood flow in both a healthy subject and a pediatric patient with a complex congenital heart defect (CHD) involving significant anatomical and hemodynamic abnormalities. We show that the image-based framework produces pressure results that agree reasonably well with catheterization measurements, and that the simulated flow fields qualitatively match the 4D-Flow MRI data of the CHD patient. Compared with 4D-Flow MRI, the simulations demonstrate strong qualitative agreement in the timing and orientation of diastolic inflow jets, systolic ejection into the aorta and pulmonary artery, and flow features associated with the narrowed pulmonary artery and the VSD. At the same time, the simulations provide a higher-resolution visualization of flow structures that are partially obscured in MRI by noise and limited spatiotemporal resolution. Finally, comparison between the healthy and CHD cases highlights differences in diastolic flow organization and ventricular energetics, including higher normalized viscous dissipation in the CHD patient.

\section{Methods}

\subsection{Patient Data}
\subsubsection{Adult patient with normal anatomy} Electrocardiogram (ECG)–gated computed tomography angiography (CTA) was obtained from a 50-year-old male patient with mild coronary artery disease. The CTA images contain 3D volumetric images at 10 phases of the cardiac cycle, sampled at every 10\% of the RR interval. The CTA volumes have $512\times512\times213$ voxels, and each voxel’s resolution is $0.39\times0.39\times0.7 \text{ mm}^3$. Segmentations of the cardiac structures, including the four chambers, aorta, and pulmonary arteries, are automatically created for all phases using a previously developed machine learning (ML) algorithm ~\cite{kong_learning_2023}. 

\subsubsection{Pediatric patient with CHD} The patient was a 15-month-old female (10.4 kg, height 81.5 cm, body surface area 0.47 m²) with congenitally corrected transposition of the great arteries (S,L,L), a large outlet perimembranous ventricular septal defect, atrial septal defect, and pulmonary valve obstruction. Namely, the morphological left ventricle was connected to the pulmonary artery through a narrowed pulmonary valve, and the morphological right ventricle was connected to the aorta, with defects (holes) in the septum between both the ventricles and atria. Pressures measured during cardiac catheterization are displayed in Table \ref{tab:cath_pressures}. 
Imaging confirmed a 51–53 mmHg pressure gradient across the pulmonary valve and narrowing of the branch pulmonary arteries, while pulmonary veins were unobstructed.

\begin{table}[h]
\centering
\caption{Cardiac catheterization pressure measurements for the CHD patient (mmHg). Ventricular and arterial pressures are reported as systolic/diastolic values. Atrial pressure represents the mean pressure.}
\label{tab:cath_pressures}
\begin{tabular}{lcccccc}
\hline
 & LV & RV & Atria & Aorta & Main PA & Branch PA \\
\hline
Pressure & 80/9 & 80/9 & 9 & 75/40 & 27/11 & 16--18/11 \\
\hline
\end{tabular}
\end{table}

Cardiac MRI with 4D-Flow imaging was performed on a 1.5 T scanner under general anesthesia using a custom protocol. The study included anatomical cine imaging (long- and short-axis views) and 4D-Flow MRI with  intravenous administration of Feraheme (31.5 mL). Data were acquired during free breathing and post-processed on a dedicated workstation to quantify chamber volumes, ventricular function, and time-resolved 3D blood flow. Ventricular volumes and systolic function were within the normal range (left ventricular ejection fraction 64\%, right ventricular ejection fraction 55\%). Quantitative 4D-Flow analysis measured an aortic flow of 1.5 L/min (3.2 L/min/m²) and a pulmonary artery flow of 2.2 L/min (4.7 L/min/m²), corresponding to a pulmonary-to-systemic flow ratio (Qp:Qs) of 1.5. The 4D-Flow acquisition spanned 487 ms, corresponding to one cardiac cycle, and was reconstructed into 30 time frames. A pretrained ML segmentation model based on a residual U-Net ~\cite{kong_sdf4chd_2024} was applied to the magnitude images across all frames. Because the model had been trained on CT rather than 4D-Flow MRI, we manually segmented cardiac structures in five representative frames, fine-tuned the model on these ground-truth labels, and then applied the adapted model to segment the remaining frames.

\subsection{Construction of Time-Series Anatomical Models}
\subsubsection{Creation of the starting meshes}

From ML-assisted segmentation of the cardiac structures, we generated surface and volume meshes of the four-chamber fluid domain for both healthy and CHD cases. The computational domain included the ventricles, the left atrium with pulmonary veins, the right atrium with superior and inferior vena cava, the aorta, and the pulmonary arteries (PA). Simulations were initialized at approximately end-diastole, just before atrial contraction (the a-wave), when the heart is in its most relaxed state. This starting phase was determined from the ventricular and atrial volume curves.

Surface meshes were created from whole-heart segmentations using the Marching Cubes algorithm~\cite{lorensen_marching_1987}. To simplify the model and avoid explicitly including downstream branches, we truncated the pulmonary veins and vena cava to retain only the segments adjacent to the atria, the pulmonary arteries to retain only the sections immediately after the left and right PA branches, and the aorta to retain only the segment before the arch. These truncations ensured that the model captured the immediate major vessel structures without extending into additional downstream branches. Surface mesh truncation and boundary face tagging were performed in SimVascular~\cite{updegrove_simvascular_2017}. Wall boundaries, where Dirichlet boundary conditions on mesh displacements were prescribed, and inlet/outlet caps, where Neumann boundary conditions were applied and coupled to a lumped parameter network model of the circulation, were explicitly defined. 

\begin{figure}[h!]
    \centering
    \begin{subfigure}[b]{\textwidth}
        \centering
        \includegraphics[width=\textwidth]{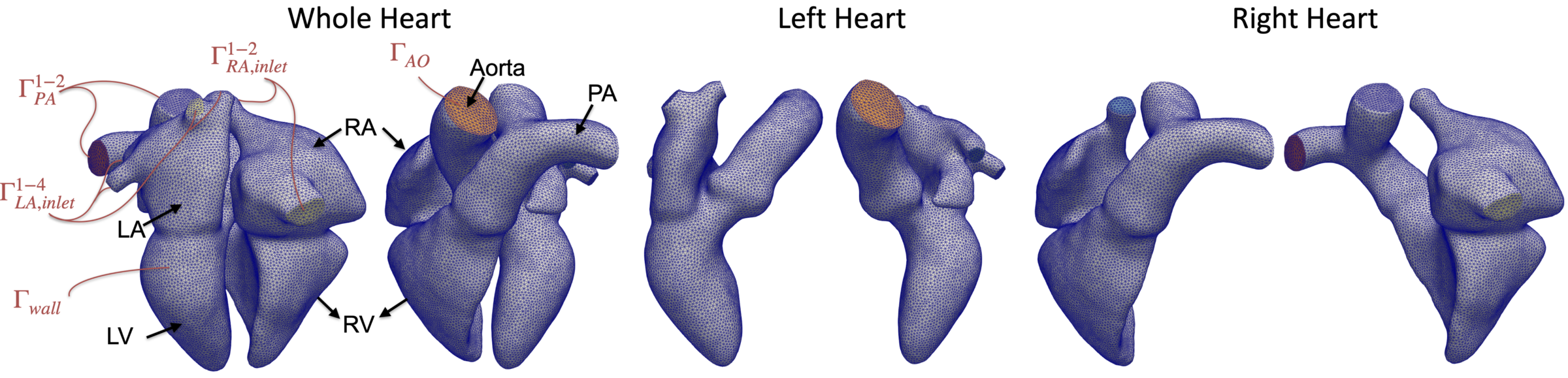}
        \caption{Healthy Patient.  }
        \label{fig:healthy_mesh}
    \end{subfigure}
    \hfill 
    \begin{subfigure}[b]{\textwidth}
        \centering \includegraphics[width=\textwidth]{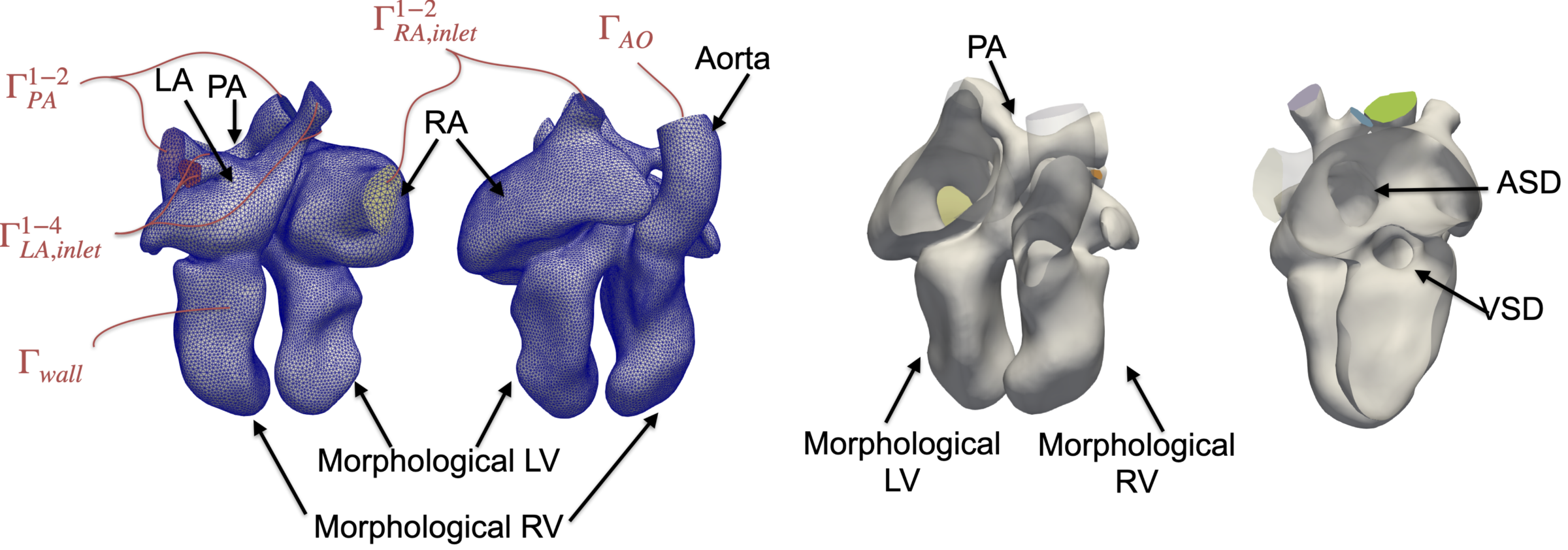}
        \caption{CHD Patient. The geometry is shown with an opacity map to reveal the PA, ASD, and VSD on the right. }
        \label{fig:vsd_mesh}
    \end{subfigure}
    \caption{Meshes constructed for the healthy patient and for the CHD patient at the beginning of the simulation (start of a-wave).}
    \label{fig:meshes}
\end{figure}

Figures \ref{fig:healthy_mesh} and \ref{fig:vsd_mesh} show the constructed surface meshes for the healthy and CHD subjects, respectively. In the CHD case,  the presence of a VSD and an ASD results in a single computational domain connecting the left and right hearts. In contrast, separate volumetric meshes are required for the right and left hearts in the healthy subject. Figure \ref{fig:vsd_mesh} also illustrates that the morphological LV is located where the RV would normally be, and vice versa, consistent with the anatomical abnormalities seen in patients with congenitally corrected transposition of the great arteries. The mitral valve and the tricuspid valve follow their embryologically linked ventricles. Thus, the mitral valve is associated with the right-sided morphological LV, and the tricuspid valve with the left-sided morphological RV. In addition, the PA is much narrower than the aorta in this CHD patient, with two further constricted regions near its bifurcation.

\subsubsection{ML-based temporal mesh reconstruction}

We adapted a neural ordinary differential equation (NODE) framework~\cite{Chen2018NeuralOD}, building on our prior shape modeling method SDF4CHD ~\cite{ kong_sdf4chd_2024}, to model diffeomorphic deformations between the baseline cardiac mesh and the anatomies at subsequent time frames. Specifically, for each time frame, we define a deformation map $\phi(\mathbf{p}, t): \mathbb{R}^3 \times \mathbb{R} \rightarrow \mathbb{R}^3$ as the solution of the NODE

\begin{equation}
\frac{\partial \phi(\mathbf{p}, t)}{\partial t} = \mathbf{v}_\theta(\phi(\mathbf{p}, t), t),
\quad \phi(\mathbf{p}, 0) = \mathbf{p}_{T_0},
\quad \phi(\mathbf{p}, T_i) = \mathbf{p}_{T_i},
\label{eq:ode}
\end{equation}

where $\mathbf{v}_\theta(\cdot, \cdot)$ is parameterized by a neural network. Existence and uniqueness of the solution are guaranteed when $\mathbf{v}_\theta$ is Lipschitz continuous. Our network consists of six fully connected layers with leaky ReLU activations ($\alpha=0.02$), which ensures differentiability and Lipschitz continuity. As a result, trajectories of mesh points do not intersect, preserving a diffeomorphic mapping between surfaces. The NODE can be integrated to yield the updated mesh coordinates:

\begin{equation}
\mathbf{p}_{T_i} = \phi(\mathbf{p}, T_i)
= \mathbf{p}_{T_0} + \int_0^{T_i} \mathbf{v}_\theta(\phi(\mathbf{p}, t), t), dt.
\label{eq:forward}
\end{equation}

We solved Equation \ref{eq:forward} using the forward Euler method with a fixed step size of 0.2 over the normalized time interval $t \in [0, 1]$.

To model the large deformations of cardiac structures across the cardiac cycle while preserving their complex anatomy, we composed four successive diffeomorphic deformation modules, each based on the NODE formulation described above. These modules progressively morphed the baseline cardiac mesh into the anatomies at subsequent time frames. To train the deformation modules, we used a composite loss function combining point matching, normal consistency, and as-rigid-as-possible (ARAP) regularization \cite{sorkine_as-rigid-as-possible_2007}:

\begin{equation}
\mathcal{L}_{\text{point}}(P_i, G_i)
= \sum_{p \in P_i} \min_{g \in G_i} | p - g |2^2
+ \sum_{g \in G_i} \min_{p \in P_i} | p - g |_2^2,
\label{eq:point_loss}
\end{equation}

\begin{equation}
\mathcal{L}_{\text{normal}}(P_i, G_i)
= \sum_{p \in P_i}
\sum_{g = \arg\min_{g \in G_i} | p - g |_2}
| (p_1 - p) \times (p_2 - p) - n_g |_2^2,
\label{eq:normal_loss}
\end{equation}

\begin{equation}
\mathcal{L}_{\text{arap}}
= \sum_{i=1}^n \sum_{j \in N(i)} w_{ij}
| (p_i' - p_j') - R_i (p_i - p_j) |_2,
\label{eq:arap_loss}
\end{equation}

Equation \ref{eq:point_loss} defines the point loss, which minimizes the surface-to-surface distance between the baseline cardiac chamber mesh and the corresponding target chamber meshes at later time frames. Here, $P_i$ denotes the predicted mesh at time $T_i$, and $G_i$ denotes the ground-truth mesh generated from segmentation. Both the starting mesh and the ground-truth meshes used during optimization were constructed from segmentation masks using the Marching Cubes algorithm. 

Because machine learning–derived segmentations can be noisy and sometimes truncate smaller vessels, a direct NODE-based deformation may otherwise collapse into singularities (e.g., a “sink” in the deformation field). To mitigate this and to preserve anatomical plausibility, we introduced two additional constraints. First, a normal consistency loss, defined in Equation \ref{eq:normal_loss}, was applied to enforce smooth orientation of surface normals and thereby discourage folding or inversion of surfaces. For a vertex $p \in P_i$, the nearest neighbor in $G_i$ is denoted by $g \in G_i$, and $n_g$ is the ground-truth surface normal at $g$. The vectors $(p_1 - p)$ and $(p_2 - p)$ are edges incident to $p$ and are used to approximate its local surface normal. Second, we incorporated the ARAP regularization term, defined in Equation \ref{eq:arap_loss}, to penalize excessive non-rigid deformations and preserve the local geometric integrity of the baseline mesh. These constraints ensured that the deformations maintained both the surface quality and the enclosed volume fidelity of the cardiac structures. In the ARAP loss, $p_i$ and $p_j$ are neighboring vertices in the reference mesh, $p_i'$ and $p_j'$ are their deformed positions, $N(i)$ denotes the set of neighbors of vertex $i$, $w_{ij}$ is a cotangent weight that balances the contributions of edges, and $R_i$ is the local optimal rotation matrix that best preserves the edge lengths around vertex $i$.

The total loss was defined as the geometric weighted mean of the three individual loss terms, 
\begin{equation}
\mathcal{L}_{\text{total}}
= \mathcal{L}_{\text{point}}^{\lambda_{\text{point}}}
\cdot \mathcal{L}_{\text{normal}}^{\lambda_{\text{normal}}}
\cdot \mathcal{L}_{\text{arap}}^{\lambda_{\text{arap}}}.
\label{eq:loss_total}
\end{equation}

Using the geometric mean allows the hyperparameters to be constrained within the range $[0,1]$, while reducing sensitivity to differences in the relative scale (magnitude) of the individual losses. To explore suitable weightings, we randomly sampled 30 sets of weights from a Dirichlet distribution with concentration parameters $\alpha = [1,1,1]$, which generates uniformly distributed triplets of non-negative values that sum to 1. From this search, we identified a set of weights, $\lambda_{\text{point}} = 0.3$, $\lambda_{\text{normal}} = 0.4$, and $\lambda_{\text{arap}} = 0.3$, that provided a good balance between registration accuracy and mesh quality.

\begin{figure}
    \centering
    \includegraphics[width=\linewidth]{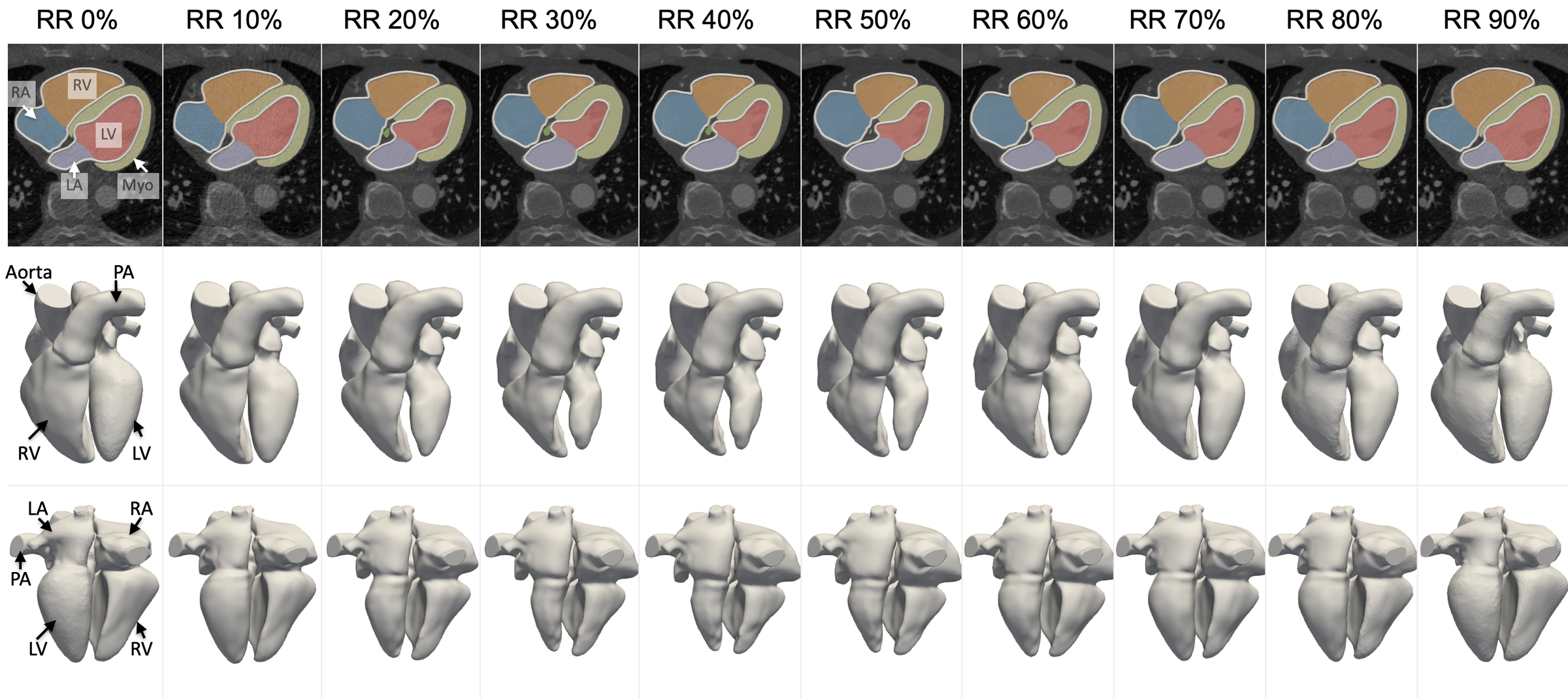}
    \caption{Reconstruction of time-series whole-heart anatomical models from imaging data for the healthy patient. (Top) Representative frames from ECG-gated CTA with automated segmentation of cardiac structures. (Middle and bottom) Corresponding 3D surface reconstructions at each phase of the cardiac cycle.}
    \label{fig:healthy_seg}
\end{figure}

\begin{figure}
    \centering
    \includegraphics[width=\linewidth]{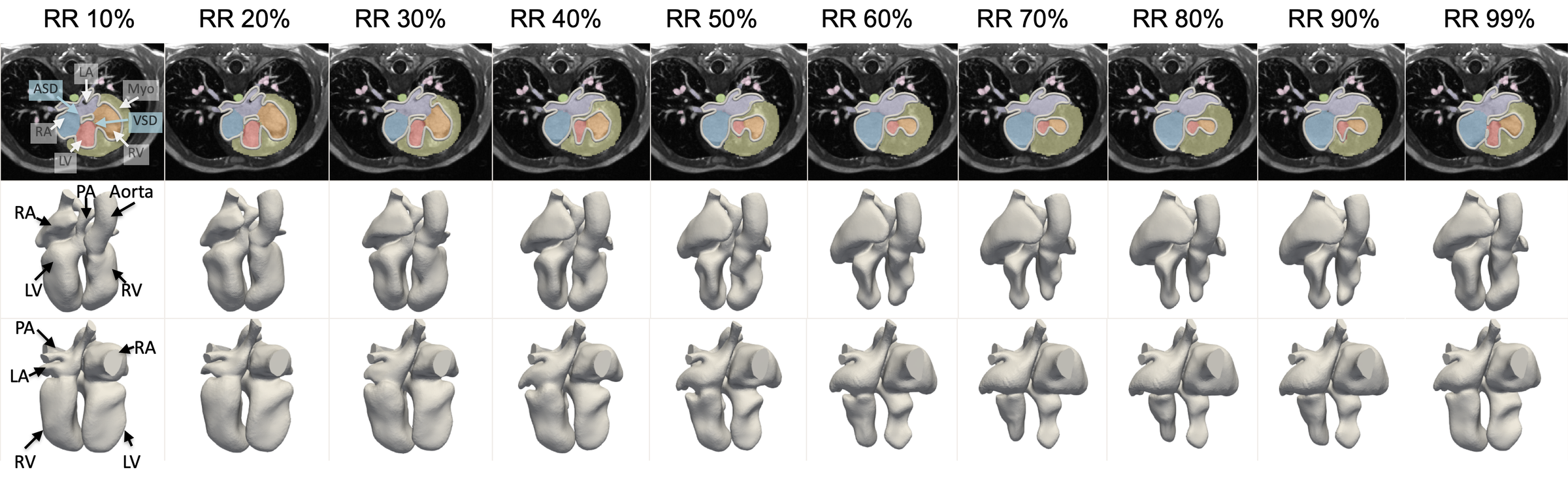}
    \caption{Reconstruction of time-series whole-heart anatomical models from imaging data for the CHD patient. (Top) Representative frames from cardiac MRI volumes with semi-automated segmentation of cardiac structures. (Middle and bottom) Corresponding 3D surface reconstructions at 10 uniformly distributed phases of the cardiac cycle. The RV and LV labels refer to the morphological RV or LV. }
    \label{fig:vsd_seg}
\end{figure}

Figures \ref{fig:healthy_seg} and \ref{fig:vsd_seg} illustrate the overall process of reconstructing time-series anatomical models from imaging data for the healthy subject and the pediatric patient with CHD, respectively. Starting from ECG-gated CTA volumes (for healthy subject) or cardiac MRI volumes (for CHD patient), cardiac chambers and great vessels were automatically segmented at multiple cardiac phases using our ML-based segmentation framework\cite{kong_learning_2023,kong_sdf4chd_2024}. From these segmentation masks, surface meshes were generated using the Marching Cubes algorithm \cite{lorensen_marching_1987}. The baseline (end-diastolic) mesh was then propagated through time using the neural ODE–based deformation model described above, resulting in temporally consistent 3D reconstructions of the heart across the cardiac cycle. The reconstructed geometries captured physiologic chamber deformation and valve-plane motion, while preserving mesh topology and smoothness throughout all phases.

\subsubsection{Creation of valve anatomies}
\begin{figure}
    \centering
    \includegraphics[width=0.5\linewidth]{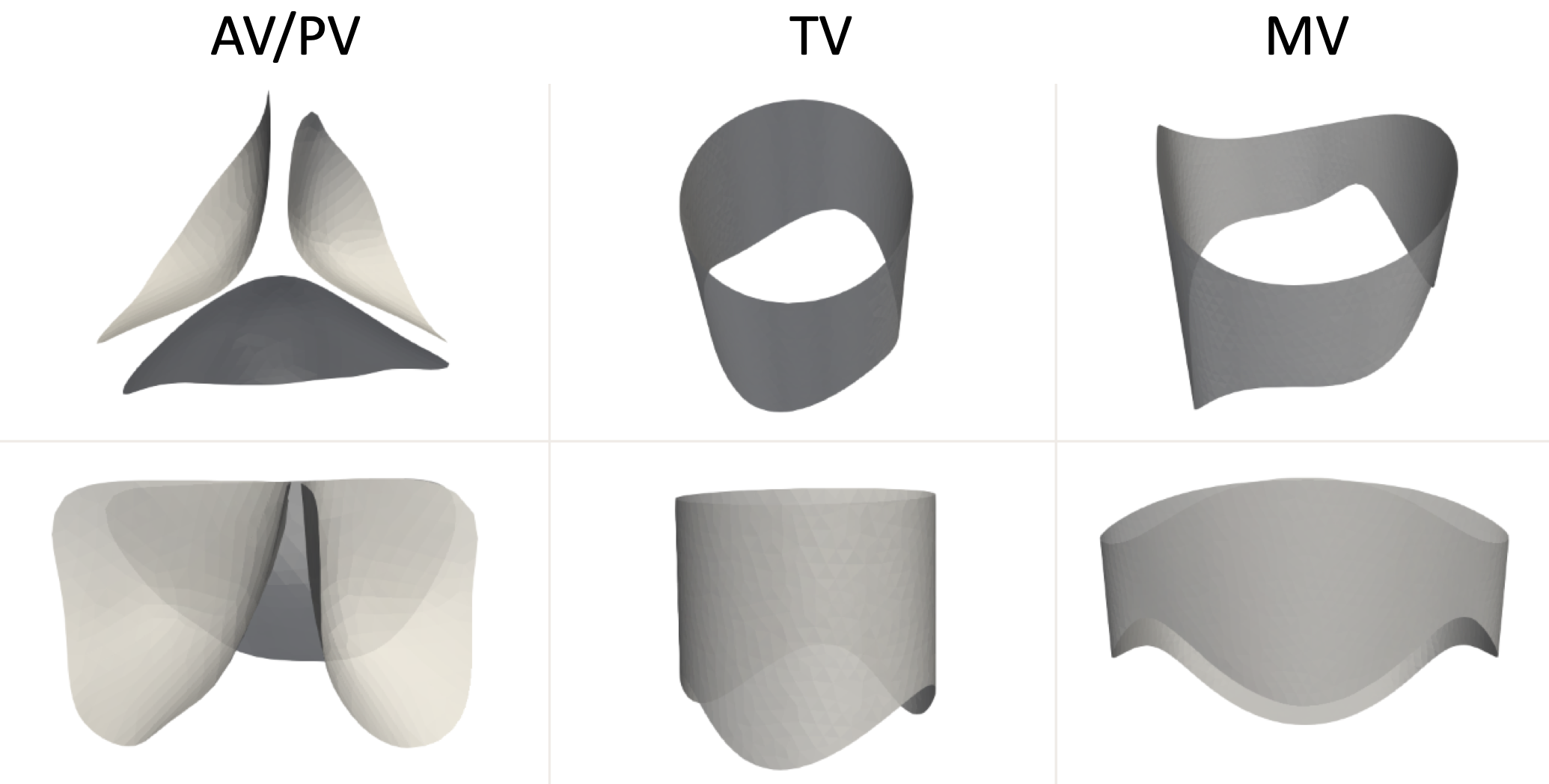}
    \caption{Template valve anatomies prior to morphing into patient-specific geometries.}
    \label{fig:valve_template}
\end{figure}

Since patient-specific valve anatomies are difficult to directly reconstruct from CT or MRI data, we adapted template anatomies of the aortic valve (AV), pulmonary valve (PV), mitral valve (MV), and tricuspid valve (TV) to fit within each patient’s cardiac geometry. The AV and PV templates were derived from manually segmented patient aortic valve anatomies curated as part of the svFSI test case~\cite{noauthor_journal_nodate}. The MV template was adapted from de Oliveira \emph{et al.}.~\cite{de_oliveira_toolbox_2021}, where the annular boundary was sampled from a porcine mitral valve mesh, and landmark points and relevant boundaries were selected to parameterize the leaflets using polynomial fitting. For the TV, we constructed an idealized template by parameterizing the free edge with sinusoidal functions to generate three leaflet cusps. Figure \ref{fig:valve_template} displays the template valve anatomies.  

To morph the template into patients' cardiac anatomies, a manually determined affine transformation matrix was first applied to position each valve within the corresponding patient-specific annulus or, for the semilunar valves, within the aortic or pulmonary root anatomy. Next, we applied a Thin Plate Spline (TPS) transform, a nonlinear warp defined by a set of source and target landmarks, to further morph the valve annulus onto the surface meshes. For the AV and PV, only the basal edge of the valve leaflets was morphed, while for the MV and TV both the annulus and the free edges were morphed. Landmark points for the annulus and basal edges were manually selected on the surface meshes, while landmarks for the free edges were initially taken from the template valves after affine transformation and adjusted to preserve the leaflet lengths. The MV and TV were modeled in the open configuration, whereas the AV and PV were modeled in the closed configuration. 

\subsection{Moving-domain CFD of cardiac flow with heart valves as resistive implicit immersed surfaces.}
\subsubsection{Moving domain CFD of cardiac flow}
We simulate blood flow using the ALE formulation, in which the cardiac chambers and great vessels are represented by moving, domain-conformal surface meshes, and the blood volume is discretized as a deforming volume mesh. The ALE method is implemented in svFSI, and details of the method can be found in Vedula \emph{et al.} ~\cite{vedula_method_2017}. Briefly, we employ linear continuous (P1-P1) finite elements for velocity and pressure, stabilized using the variational multiscale (VMS) method~\cite{liu_unified_2018}, which incorporates pressure-stabilizing/Petrov–Galerkin (PSPG) stabilization to address the instabilities of the standard Galerkin method. Temporal discretization is performed with the second-order generalized-$\alpha$ method~\cite{liu_unified_2018}, and nonlinear terms are linearized via a modified Newton–Raphson scheme \cite{esmaily-moghadam_bi-partitioned_2015}. Mesh motion is solved using a linear elastostatic model augmented with Jacobian-based stiffening, coupled to the fluid equations through a block-iterative quasi-direct scheme where mesh motion lags by one iteration~\cite{vedula_method_2017}. The resulting sparse linear systems are solved using the generalized minimal residual method (GMRES) \cite{saad_gmres_1986} with Jacobi preconditioning. The solver is parallelized with MPI for large-scale cardiovascular simulations and incorporates backflow stabilization to avoid unphysiological flow reversal at the Neumann boundary surfaces.
\begin{equation}
\label{eq:NS_ALE}
\left\{
\begin{alignedat}{2}
&\rho\!\left(\frac{\partial \mathbf{u}}{\partial t}
+ \bigl((\mathbf{u}-\mathbf{u}_{\mathrm{ALE}})\!\cdot\!\nabla\bigr)\mathbf{u}\right)
- \nabla\!\cdot\!\bigl( \mu(\nabla\mathbf{u}+\nabla\mathbf{u}^{T}) \bigr)
+ \nabla p \;=\; \mathbf{0},
\quad &&\text{in }\Omega_t,\\[2pt]
&\nabla\!\cdot\!\mathbf{u} \;=\; 0,
\quad &&\text{in }\Omega_t,
\end{alignedat}
\right.
\end{equation}

\begin{equation}
\label{eq:BC_fluid}
\left\{
\begin{alignedat}{2}
&\bigl(-p\mathbf{I}+\mu(\nabla\mathbf{u}+\nabla\mathbf{u}^{T})\bigr)\mathbf{n} \;=\; \bar{\mathbf{t}},
\quad &&\text{on } \Gamma_{N}(t)\ \text{(caps)},\\[2pt]
&\mathbf{u} \;=\; \mathbf{u}_{\mathrm{ALE}},
\quad &&\text{on } \Gamma_{w}(t)\ \text{(moving wall)}.
\end{alignedat}
\right.
\end{equation}

\subsubsection{Heart valves as implicit resistive immersed surfaces (RIS)}
\noindent
We used a collection of immersed surfaces 
\(\{\Gamma_k\}_{k \in \mathcal{I}_v}\) within the fluid domain \(\Omega_t\), to represent the cardiac valves ~\cite{zingaro_modeling_2023,davey_simulating_2024}. 
Each surface \(\Gamma_k\) is implicitly defined by a signed distance function 
\(\varphi_k:\Omega_t \times (0,T)\to \mathbb{R}\), 
so that
\[
  \Gamma_k \;=\; \bigl\{\mathbf{x}\in \Omega_t : \varphi_k(\mathbf{x})=0 \bigr\}, 
  \quad \text{for } k\in \mathcal{I}_v.
\]

Instead of imposing explicit boundary conditions on \(\Gamma_k\), the resistive immersed surface method introduces a volumetric penalization in the momentum equation. For each valve, the added force density is 
\[  \mathbf{f}_{RIS, k} = \frac{R_k}{\epsilon_k} \delta_{\Gamma_k}\bigl(\varphi_k(\mathbf{x})\bigr) (\mathbf{u} - \mathbf{u}_{ALE}), \]

where \(R_k\) is a resistance coefficient and \(\epsilon_k\) is typically chosen on the order of the mesh size
and can be interpreted as half the effective thickness of the valve.

To localize this penalization within a narrow region around \(\Gamma_k\), we employ a smoothed Dirac delta function:
\[
  \delta_{\Gamma_k}\bigl(\varphi_k(\mathbf{x})\bigr)
  \;=\;
  \begin{cases}
    \displaystyle
    \frac{1 + \cos\!\Bigl(\pi\,\frac{\varphi_k(\mathbf{x})}{\varepsilon_k}\Bigr)}{2\,\varepsilon_k},
    & \text{if } \bigl|\varphi_k(\mathbf{x})\bigr|\le \varepsilon_k,\\[6pt]
    0,
    & \text{if } \bigl|\varphi_k(\mathbf{x})\bigr|> \varepsilon_k.
  \end{cases}
\]
This smoothed delta function ensures that the resistive force is only applied 
in the vicinity of the immersed valve surface. With this RIS forcing term, the fluid momentum equation becomes: 

\begin{equation}
\label{eq:NS_ALE_RIIS}
\left\{
\begin{alignedat}{3}
\rho \Biggl(\frac{\partial \mathbf{u}}{\partial t} 
+ \bigl((\mathbf{u}-\mathbf{u}_{\text{ALE}})\cdot\nabla\bigr)\mathbf{u}\Biggr)
- \nabla \cdot \Bigl(\mu(\nabla \mathbf{u} + \nabla \mathbf{u}^T)\Bigr)
+ \nabla p + \sum_{k \in \mathcal{K}} \frac{R_k}{\varepsilon_k}\,\delta_{\Gamma_k}(\varphi_k)\,
\Bigl(\mathbf{u} - \mathbf{u}_{\text{ALE}} \Bigr)
&= 0, \quad &\text{in } \Omega_t,\\[2pt]
\nabla \cdot \mathbf{u}
&= 0, \quad &\text{in } \Omega_t.
\end{alignedat}
\right.
\end{equation}

The signed distance function $\varphi_k$ is computed on the reference fluid domain. When a valve changes its status, we apply a prescribed motion from open to closed or from closed to open and recompute the signed distance function. The prescribed motions are extracted from valve mechanical simulations described in the next section (Section \ref{sec:valve_contact}). The computations of the signed distance fields are performed in the reference configuration without considering the velocity of the fluid mesh. The velocity of the fluid mesh subsequently morphs both the fluid domain and the embedded implicit valve surfaces. To reduce computational cost, the signed distance function is only recomputed when the valve status changes, rather than at every time step.

The valve status change is triggered by two different criteria, depending on the current valve state ~\cite{this_augmented_2020, zingaro_modeling_2023}. If a valve is currently closed, we evaluate the pressure upstream and downstream of the valve. For aortic or pulmonary valves, the valve opens at the next time step if the pressure on the ventricular surface (upstream) exceeds the pressure on the arterial surface (downstream). Similarly, for mitral and tricuspid valves (atrioventricular valves), the valve opens if the pressure on the atrial surface (upstream) exceeds the pressure on the ventricular surface (downstream). In practice, the average pressure is computed over regions defined by the signed distance field: regions with $\psi_k(\mathbf{x}) > 2.5\epsilon_k$ are considered to lie in the upstream region, and regions with $\psi_k(\mathbf{x}) < -2.5\epsilon_k$ are considered to lie in the downstream region.

When a valve is currently open, we check for backflow, which may trigger closure. If an average backflow is detected in the fluid region enclosed by the valve, the valve begins its closing motion at the next time step. The region used to compute the average flow rate is defined as $\psi_k(\mathbf{x}) < -\epsilon_k$, which corresponds to the region inside the valve surface. Once a status change is triggered, the valve undergoes the prescribed opening or closing motion until the transition is completed.

\subsection{Valve mechanics simulations with contact}
\label{sec:valve_contact}
To account for physiologically realistic valve dynamics, we performed separate simulations of valve opening and closing driven by transvalvular pressure differences for each valve and each patient. The resulting valve kinematics were then extracted and prescribed within the ALE–RIS simulations when a change in valve state was triggered. Valve mechanics were modeled using shell structures within svFSI\cite{zhu_svfsi_2022}. The mitral and tricuspid valve leaflets were modeled as thin shells using a Lee–Sacks constitutive model. The strain energy function follows the formulation in \cite{lee2014inversemitralvalve}, and the material parameters $(a, a_0, b_1, b_2, \mu_0) = (2.0, 2.9684, 2.661, 1.0, 1.0)$ were used for both cases. The shell thickness was set to 0.4 mm. A transvalvular pressure was ramped up to 80 mmHg, or until the leaflets reached a fully closed configuration. The leaflet annulus was fixed in place, and a distributed traction was applied along the free edges to represent forces from the chordae tendineae during the simulation. The AV and PV were modeled using a Neo-Hookean constitutive model with an elasticity modulus of 0.5 MPa and a Poisson ratio of 0.45. The shell thickness was also set to 0.4 mm. The transvalvular pressure was gradually increased until the leaflets reached a fully open configuration. 

Leaflet contact was treated using the penalty-based contact formulation to prevent interpenetration between opposing structural surfaces. A brief overview is provided here. Assume that there are two contacting surfaces $S_1$ and $S_2$. A discrete set of candidate contact points 
$\{\mathbf{x}_{1,i}\}_{i=1}^{N_c}$ is defined on one leaflet surface $S_1$.  
For each point $\mathbf{x}_{1,i}$, a closest-point projection onto the opposing 
surface $S_2$ identifies the corresponding point $\mathbf{x}_{2,i}$, together with 
the outward unit normal $\mathbf{n}_{2,i}$ at $\mathbf{x}_{2,i}$.  
Following~\cite{kamensky2015immersogeometric}, the signed penetration distance is 
computed as
\begin{equation}
    d_i = (\mathbf{x}_{2,i} - \mathbf{x}_{1,i}) \cdot \mathbf{n}_{2,i},
    \label{eq:penetration}
\end{equation}
where $d_i > -h$ indicates that the two surfaces are closer than the admissible 
minimum gap $h \ge 0$, and $d_i > 0$ corresponds to interpenetration.  
The parameter $h$ provides a smooth transition into contact and avoids numerical 
instabilities that may arise from an abrupt onset of the penalty response.

To enforce non-penetration, we follow the penalty formulation used in 
\cite{kamensky2015immersogeometric}, in which the penetration distance is mapped 
to a scalar penalty magnitude by
\begin{equation}
P_k(d) =
\begin{cases}
\displaystyle \frac{k}{2h}(d+h)^2, & d \in (-h,0), \\[8pt]
\displaystyle \frac{kh}{2} + k\,d, & d \ge 0, \\[6pt]
0, & d \le -h,
\end{cases}
\label{eq:Pk_penalty}
\tag{59}
\end{equation}
where $k > 0$ is the penalty stiffness.  
The quadratic branch ensures a smooth increase in force as the surfaces approach 
one another, while the linear branch prevents excessive stiffness for larger values 
of penetration.

The resulting contact forces applied to each point pair 
$(\mathbf{x}_{1,i},\mathbf{x}_{2,i})$ are given by
\begin{equation}
    \mathbf{f}_{1,i} = -\, w_i\, P_k(d_i)\,\mathbf{n}_{2,i}, 
    \qquad 
    \mathbf{f}_{2,i} = -\mathbf{f}_{1,i},
    \label{eq:contact_forces}
\end{equation}
where $w_i$ is the quadrature weight associated with point $\mathbf{x}_{1,i}$.  
These forces act along the line of separation between the point pair and are equal 
and opposite, ensuring conservation of linear momentum.  
Because the direction of action is aligned with the normal at $\mathbf{x}_{2,i}$, 
the formulation also conserves angular momentum, as described in 
\cite{kamensky2015immersogeometric}.

The discrete forces $\mathbf{f}_{1,i}$ and $\mathbf{f}_{2,i}$ are accumulated into the 
global structural force vector in the standard finite element fashion.  
Since the method operates directly on point pairs, it naturally accommodates 
large deformations, sliding contact, and repeated separation and re-contact, 
without requiring a surface-to-surface mapping or additional constraint variables. Overall, this point-to-point penalty strategy provides a robust and computationally 
simple framework for representing leaflet contact while maintaining compatibility 
with the standard finite element formulation used in this work.

\subsection{0D lumped parameter networks of circulation}

\begin{figure}[h]
    \centering
    \includegraphics[width=\linewidth]{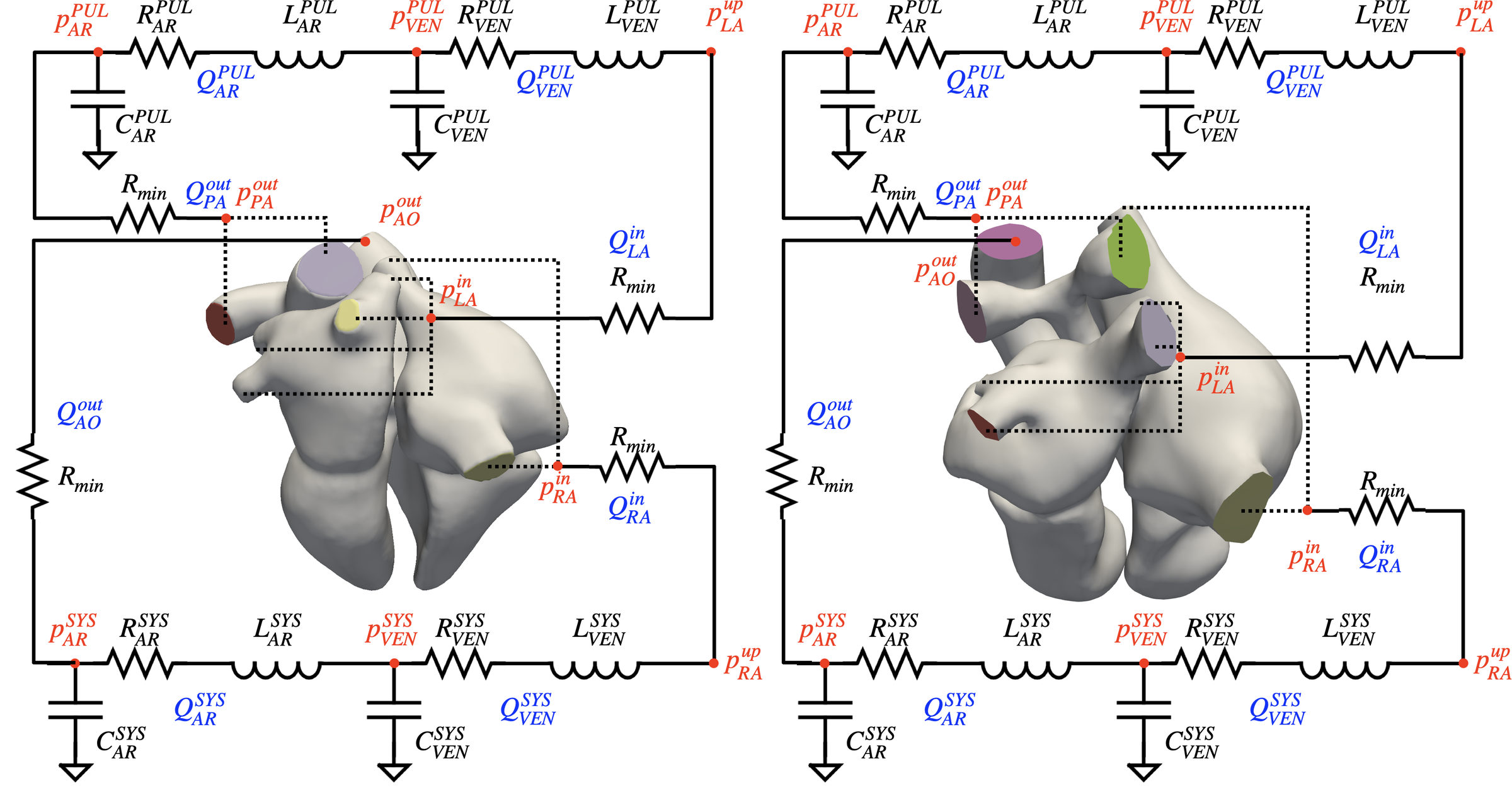}
    \caption{Closed-loop coupling between the 3D RIS–ALE cardiac flow model and the LPN for the healthy (left) and CHD (right) cases.}
    \label{fig:LPN}
\end{figure}

To model physiologically realistic pressure dynamics at the inlets and outlets, we coupled our 3D CFD simulation with a 0D lumped parameter network (LPN) representing the systemic and pulmonary circulations, adapted from~\cite{regazzoni_cardiac_2022,brown_modular_2024}. The closed-loop LPN, illustrated in Fig.~\ref{fig:LPN}, consists of capacitor-resistor-inductor (C-R-L) assemblies representing four compartments: systemic arteries, systemic veins, pulmonary arteries, and pulmonary veins.

We employ a modular, fully implicit 3D–0D coupling framework recently developed in~\cite{brown_modular_2024}, enabling bidirectional exchange between the 3D flow model and the 0D circulation model. Specifically, flow rates computed at the 3D inlets and outlets are passed to the 0D model as input, while the resulting pressures at the LPN nodes ( $p_{\text{LA}}^{in}$, $p_{\text{RA}}^{in}$, $p_{\text{AO}}^{out}$, $p_{\text{PA}}^{out}$) are fed back and imposed as Neumann boundary conditions on the 3D domain (Fig.~\ref{fig:LPN}).

The 0D ODE system is integrated using an explicit fourth-order Runge–Kutta (RK4) method. Coupling is performed within each Newton iteration of the 3D solver, during which flow and pressure are exchanged between 0D and 3D models.

Specifically, the ODEs governing the LPN are as follows:
\begin{subequations}
\label{eq:system}
\begin{align}
\frac{d\,p_{AR}^{SYS}}{dt} 
  &= \frac{1}{C_{AR}^{SYS}}\bigl(Q_{AO} - Q_{AR}^{SYS}\bigr), \\[6pt]
\frac{d\,p_{VEN}^{SYS}}{dt}
  &= \frac{1}{C_{VEN}^{SYS}}\bigl(Q_{AR}^{SYS} - Q_{VEN}^{SYS}\bigr), \\[6pt]
\frac{d\,p_{AR}^{PUL}}{dt}
  &= \frac{1}{C_{AR}^{PUL}}\bigl(Q_{PA} - Q_{AR}^{PUL}\bigr), \\[6pt]
\frac{d\,p_{VEN}^{PUL}}{dt}
  &= \frac{1}{C_{VEN}^{PUL}}\bigl(Q_{AR}^{PUL} - Q_{VEN}^{PUL}\bigr), \\[6pt]
\frac{d\,Q_{AR}^{SYS}}{dt}
  &= -\frac{R_{AR}^{SYS}}{L_{AR}^{SYS}}\,Q_{AR}^{SYS}
     -\frac{p_{VEN}^{SYS} - p_{AR}^{SYS}}{L_{AR}^{SYS}}, \\[6pt]
\frac{d\,Q_{AR}^{PUL}}{dt}
  &= -\frac{R_{AR}^{PUL}}{L_{AR}^{PUL}}\,Q_{AR}^{PUL}
     -\frac{p_{VEN}^{PUL} - p_{AR}^{PUL}}{L_{AR}^{PUL}}.
\end{align}
\end{subequations}

This system involves six unknown, time-dependent variables: systemic arterial pressure $p_{AR}^{SYS}(t)$, systemic venous pressure $p_{VEN}^{SYS}(t)$, pulmonary arterial pressure $p_{AR}^{PUL}(t)$, pulmonary venous pressure $p_{VEN}^{PUL}(t)$, systemic arterial flow rate $Q_{AR}^{SYS}(t)$, and pulmonary arterial flow rate $Q_{AR}^{PUL}(t)$. Initial conditions for each variable must be specified at $t = 0$.

Flow inputs to the 0D model are derived from the 3D simulation as:
\begin{subequations}
\label{eq:inflow}
\begin{align}
Q_{AO}^{out} &= \int_{\Gamma_{AO}} \mathbf{u} \cdot \mathbf{n}_{AO}\, d\Gamma, \\
Q_{PA}^{out} &= \sum_{i=1}^{2} \int_{\Gamma_{PA}^{i}} \mathbf{u} \cdot \mathbf{n}_{PA}^{i}\, d\Gamma, \\
Q_{RA}^{in} &\equiv Q_{VEN}^{SYS} = \sum_{i=1}^{2} \int_{\Gamma_{RA}^{i}} \mathbf{u} \cdot \mathbf{n}_{RA}^{i}\, d\Gamma, \\
Q_{LA}^{in} &\equiv Q_{VEN}^{PUL} = \sum_{i=1}^{4} \int_{\Gamma_{LA}^{i}} \mathbf{u} \cdot \mathbf{n}_{LA}^{i}\, d\Gamma.
\end{align}
\end{subequations}
Here, $\mathbf{u}$ denotes the 3D velocity and $\mathbf{n}$ denotes the outward unit normal vector on each inlet or outlet surface $\Gamma$.

After solving the ODE system in Eq.~\eqref{eq:system}, the corresponding pressures at the boundaries are computed using:
\begin{subequations}
\label{eq:bc_pressure}
\begin{align}
p_{LA}^{in} &= p_{VEN}^{PUL}
  - R_{VEN}^{PUL} Q_{VEN}^{PUL}
  - L_{VEN}^{PUL} \frac{d}{dt} Q_{VEN}^{PUL}
  - R_{\min} Q_{VEN}^{PUL}, \\[6pt]
p_{AO}^{out} &= p_{AR}^{SYS} 
  + R_{\min} Q_{AO}, \\[6pt]
p_{RA}^{in} &= p_{VEN}^{SYS}
  - R_{VEN}^{SYS} Q_{VEN}^{SYS}
  - L_{VEN}^{SYS} \frac{d}{dt} Q_{VEN}^{SYS}
  - R_{\min} Q_{VEN}^{SYS}, \\[6pt]
p_{PA}^{out} &= p_{AR}^{PUL} 
  + R_{\min} Q_{PA}.
\end{align}
\end{subequations}

These pressure values are imposed as Neumann boundary conditions on the corresponding inlets and outlets of the 3D domain. For simplicity, we assume equal pressure across all pulmonary vein inlets, both vena cava inlets (SVC and IVC), and both pulmonary artery branches.

\section{Experiments and Results}

\subsection{Computational setup}

From the surface meshes, tetrahedral volume meshes were generated using TetGen (Table \ref{tab:mesh}). The thickness parameter $\epsilon_k$ for the healthy subject was approximated based on thickness values reported for adult human hearts~\cite{sahasakul_age-related_1988}, whereas for the CHD patient it was estimated by extrapolating the age-dependent thickness measurements and scaling according to the relative valve size ~\cite{sahasakul_age-related_1988,ong_biomechanics_2021}. The CFD meshes were refined near the valve regions such that  $2\epsilon_k$ value was greater than the maximum edge size, to prevent flow leakage across the valve surfaces.

\begin{table}[h!]
\caption{Summary of computational settings.}
\label{tab:mesh}
\begin{tabular}{llllllllllll}
\hline
Subject & \multicolumn{3}{l}{Mesh size {[}mm{]}} & \# of points & \# of cells & $\Delta t$    &  cycle time (s) & \multicolumn{4}{l}{$\epsilon_k$ {[}mm{]}}  \\ \cline{9-12} 
        & Min         & Avg         & Max        &        &      &            &               & MV     & AV     & TV    & PV     \\ \hline
Healthy &       0.60       &     1.79        &    4.43        & 148,317      & 864,140     & $6.896e^{-4}$ & 0.690 & 2.0    & 2.0    & 2.0   & 2.0    \\
CHD     &       0.21      &     0.99        &     1.89       & 105,783      & 594,359     & $4.276e^{-4}$ &  0.496 & 0.5    & 0.3   & 0.5   & 0.2   \\ \hline
\end{tabular}

\end{table}

\begin{figure}[h!]
    \centering
    \includegraphics[width=0.8\linewidth]{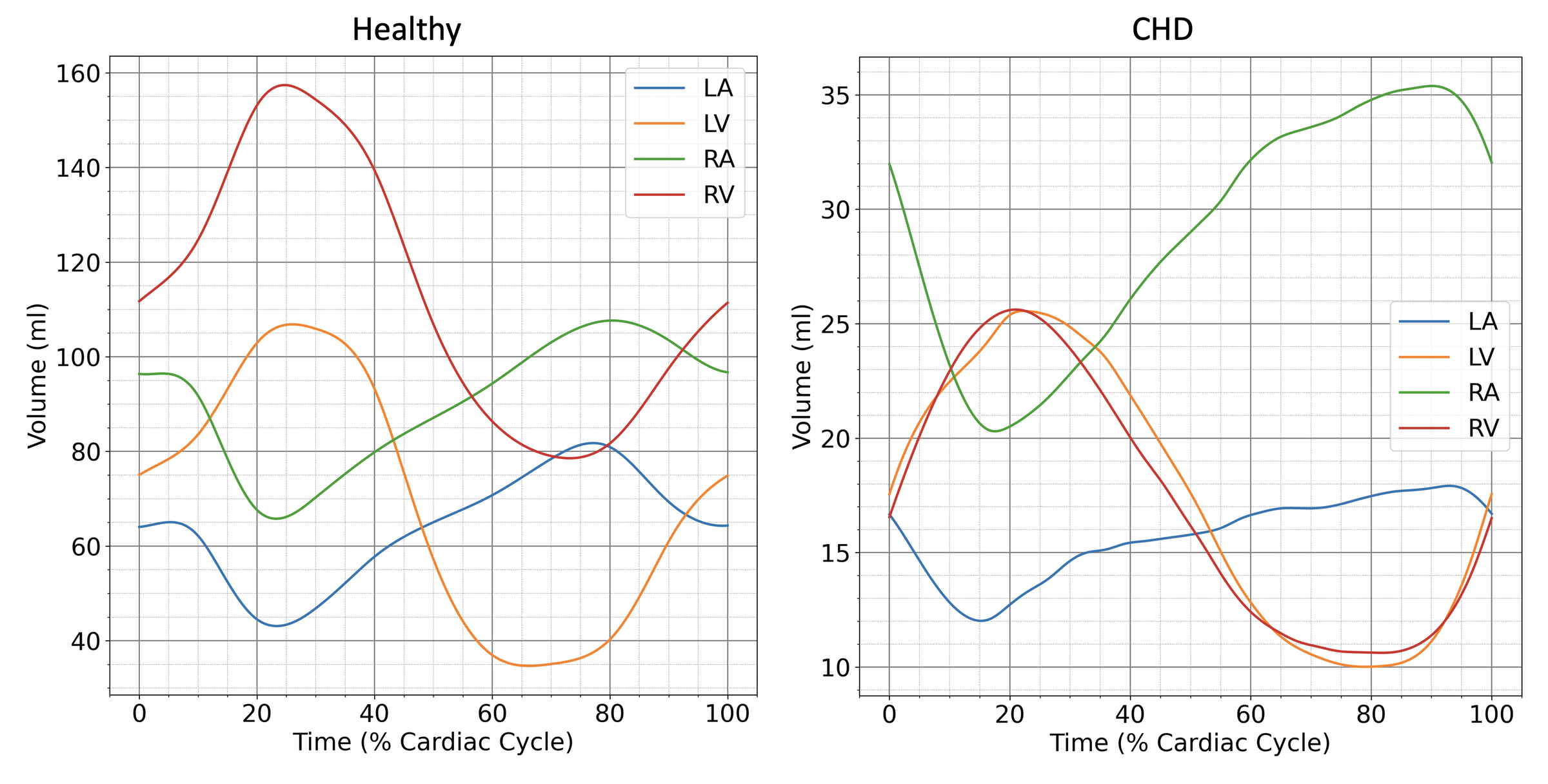}
    \caption{Cardiac chamber volume curves after applying wall motion boundary conditions obtained from image registration. Left: healthy subject; right: CHD patient.}
    \label{fig:volume_curve}
\end{figure}

Figure \ref{fig:volume_curve} shows the cardiac chamber volume curves obtained after applying the wall motion boundary conditions for both the healthy subject and the CHD patient. The curves start at the end-diastolic phase, just before atrial contraction, illustrating the reduction in atrial volume during atrial contraction, followed by the gradual volume increase during the reservoir phase (ventricular systole) and the subsequent decrease during the conduit phase (ventricular diastole) as blood flows back into the ventricles. For the ventricles, the curves capture the expected volume increase during diastole and decrease during systole. When comparing the two cases, the CHD patient exhibits similar volumes for the LV and RV, whereas in the healthy subject, the RV maintains a consistently larger volume than the left. Additionally, the CHD patient shows an enlarged RA and a relatively small LA compared with the other chambers.

\begin{figure}[h!]
    \centering
    \begin{subfigure}[b]{\textwidth}
        \centering
        \includegraphics[width=\textwidth]{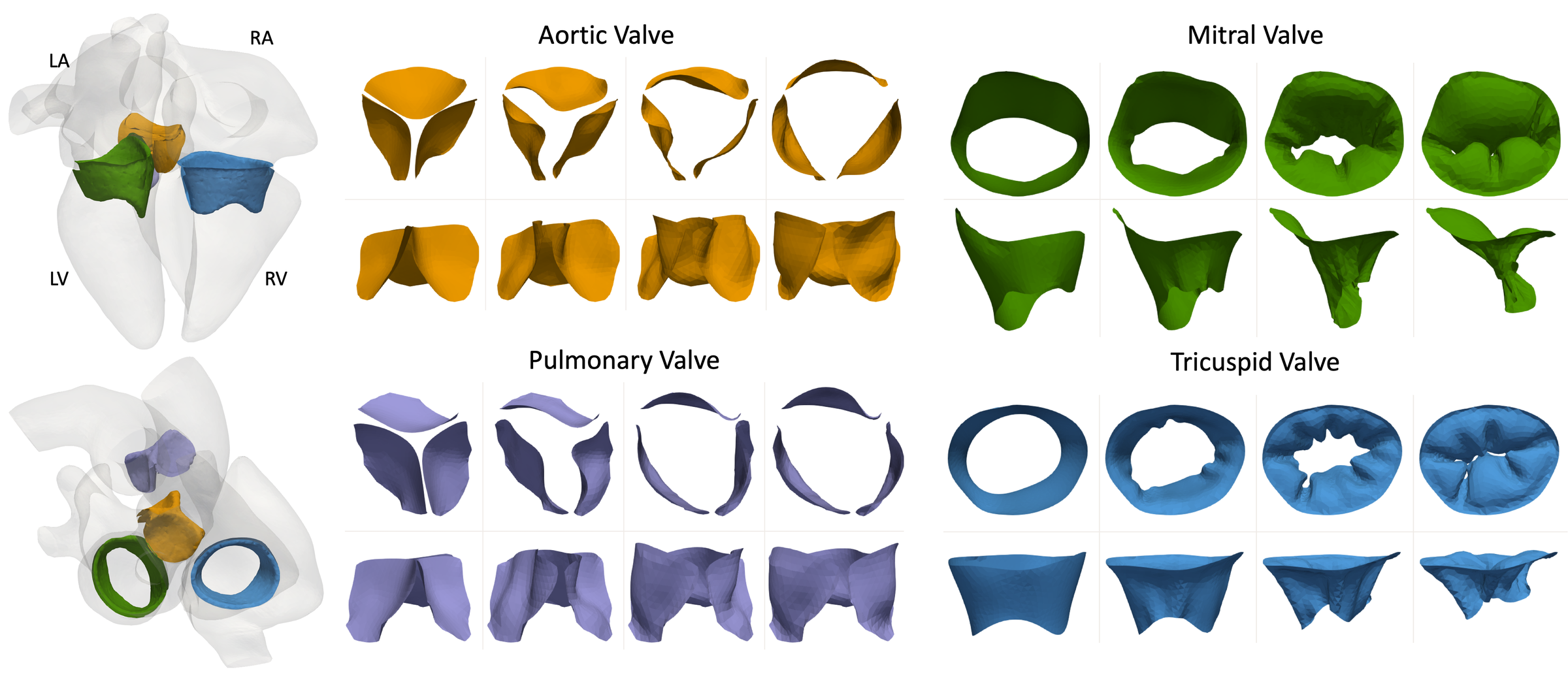}
        \caption{Healthy  }
        \label{fig:healthy_valve}
    \end{subfigure}
    \hfill 
    \begin{subfigure}[b]{\textwidth}
        \centering \includegraphics[width=\textwidth]{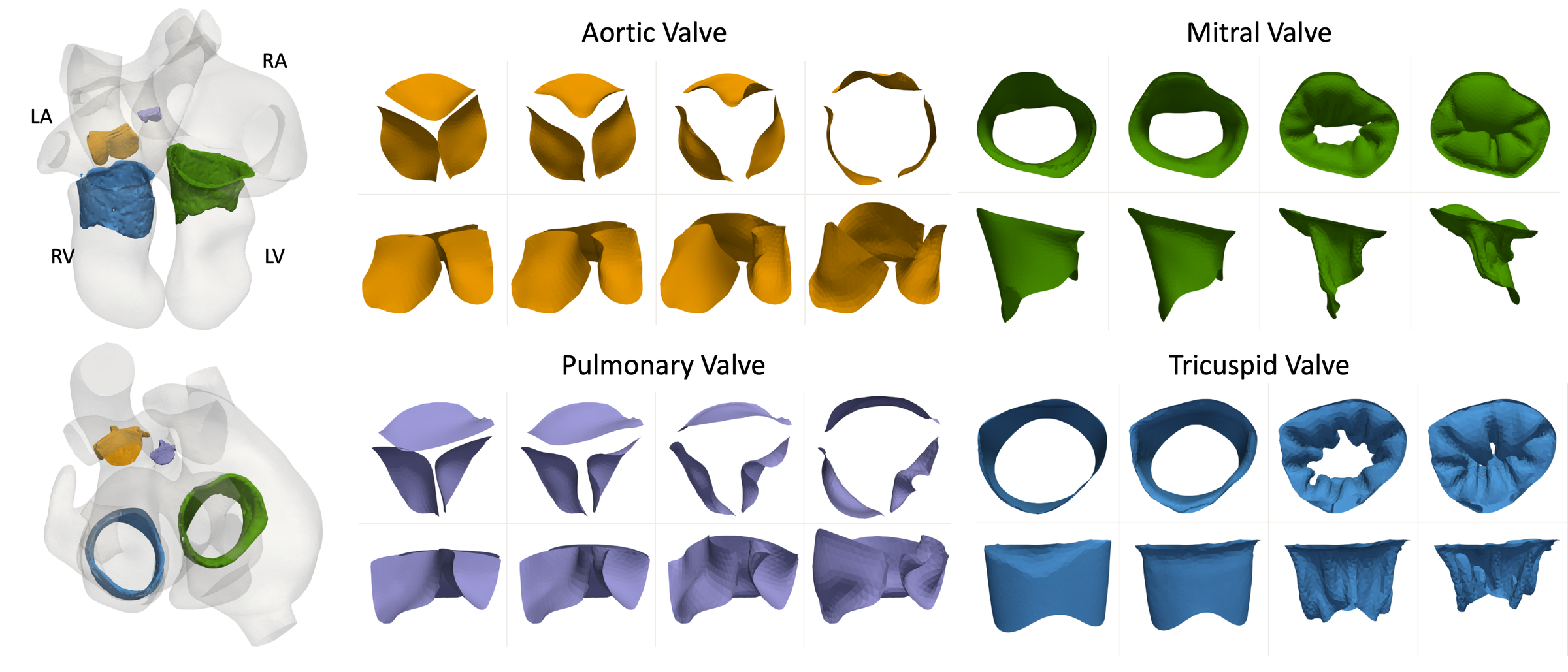}
        \caption{CHD }
        \label{fig:vsd_valve}
    \end{subfigure}
    
    \caption{Valve setup for the healthy subject and the CHD patient. Left: Iso-surface of the signed-distance function defining the valve region where the RIS resistance penalty is applied, extracted at the start of the simulation. Right: Simulated valve motion illustrating valve anatomy at different phases of opening and closing. All visualizations are shown in two different views.}
    \label{fig:valve_setup}
\end{figure}

Figure \ref{fig:valve_setup} illustrates the valve configurations and motions incorporated into the RIS simulations. The iso-surfaces of the signed-distance fields define the spatial regions where the resistive penalty is applied. The simulated valve motion represents the prescribed temporal evolution of valve resistance once the opening or closing conditions are triggered. In the shell simulations performed with svFSI, small residual gaps were observed in the closed aortic and pulmonary valves and small orifice areas in the closed mitral and tricuspid valves. These gaps, however, remain sufficiently small to prevent leakage during the CFD simulation, as the valve resistance is applied within a finite neighborhood of the valve shell surface with a specified thickness parameter.

\subsection{Simulation results for the healthy subject}

Table \ref{tab:0D_params} summarizes the parameters used in the closed-loop 0D LPN for the healthy case. These parameters were adopted from Brown \emph{et al.}  \cite{brown2025biventricular}, who optimized a similar circulatory model within a biventricular mechanics framework to match patient-specific cuff blood pressure measurements and literature-reported pressure ranges. In the coupled simulations, the 0D LPN provides Neumann boundary conditions at the outlets of the three-dimensional (3D) fluid domain, while the prescribed wall motion supplies Dirichlet boundary conditions for the mesh velocity.

\renewcommand{\arraystretch}{1.2} 
\begin{longtable}{l c c c}
\caption{Parameters of the 0D circulation model for the healthy case. } \label{tab:0D_params} \\
\hline
\textbf{Parameter} & \textbf{Symbol} & \textbf{Value} & \textbf{Units} \\
\hline
\endfirsthead

\multicolumn{4}{c}%
{\tablename\ \thetable\ -- \textit{continued from previous page}} \\
\hline
\textbf{Parameter} & \textbf{Symbol} & \textbf{Value} & \textbf{Units} \\
\hline
\endhead

\hline \multicolumn{4}{r}{\textit{Continued on next page}} \\
\endfoot

\hline
\endlastfoot
\textbf{\textit{Systemic circulation}}  \\
Systemic arterial resistance & $R_\mathrm{AR}^\mathrm{SYS}$ & \hide{0.64} $0.677 $ & mmHg s mL$^{-1}$ \\
Systemic arterial capacitance & $C_\mathrm{AR}^\mathrm{SYS}$ & \hide{1.2} $0.925 $ & mL mmHg$^{-1}$ \\
Systemic arterial inductance & $L_\mathrm{AR}^\mathrm{SYS}$ & $0.005$ & mmHg s$^2$ mL$^{-1}$ \\
Systemic venous resistance & $R_\mathrm{VEN}^\mathrm{SYS}$ & \hide{0.32} $0.064 $ & mmHg s mL$^{-1}$ \\
Systemic venous capacitance & $C_\mathrm{VEN}^\mathrm{SYS}$ & $60.0$ & mL mmHg$^{-1}$ \\
Systemic venous inductance & $\displaystyle L_\mathrm{VEN}^\mathrm{SYS}$ & $0.0005$ & mmHg s$^2$ mL$^{-1}$ \\

\\
\textbf{\textit{Pulmonary circulation}}  \\
Pulmonary arterial resistance & $R_\mathrm{AR}^\mathrm{PUL}$ & $0.032$ & mmHg s mL$^{-1}$ \\
Pulmonary arterial capacitance & $C_\mathrm{AR}^\mathrm{PUL}$ & $10.0$ & mL mmHg$^{-1}$ \\
Pulmonary arterial inductance & $L_\mathrm{AR}^\mathrm{PUL}$ & $0.0005$ & mmHg s$^2$ mL$^{-1}$ \\
Pulmonary venous resistance & $R_\mathrm{VEN}^\mathrm{PUL}$ & $0.035$ & mmHg s mL$^{-1}$ \\
Pulmonary venous capacitance & $C_\mathrm{VEN}^\mathrm{PUL}$ & $16.0$ & mL mmHg$^{-1}$ \\
Pulmonary venous inductance & $L_\mathrm{VEN}^\mathrm{PUL}$ & $0.0005$ & mmHg s$^2$ mL$^{-1}$ \\

\\
\textbf{\textit{Initial conditions}}  \\
Systemic arterial pressure & $p_\mathrm{AR}^\mathrm{SYS}(0)$ & $87.25$ & mmHg \\
Systemic venous pressure & $p_\mathrm{VEN}^\mathrm{SYS}(0)$ & \hide{37.023} $14.703 $ & mmHg \\
Pulmonary arterial pressure & $p_\mathrm{AR}^\mathrm{PUL}(0)$ & $17.73$ & mmHg \\
Pulmonary venous pressure & $p_\mathrm{VEN}^\mathrm{PUL}(0)$ & $13.83$ & mmHg \\
Systemic arterial flow & $Q_\mathrm{AR}^\mathrm{SYS}(0)$ & $110.9$ & mL s$^{-1}$ \\
Pulmonary arterial flow & $Q_\mathrm{AR}^\mathrm{PUL}(0)$ & $123.5$ & mL s$^{-1}$ \\
\hline
\end{longtable}


\begin{figure}[h!]
  \centering
  \begin{subfigure}[t]{0.49\textwidth}
    \centering
    \includegraphics[width=\linewidth]{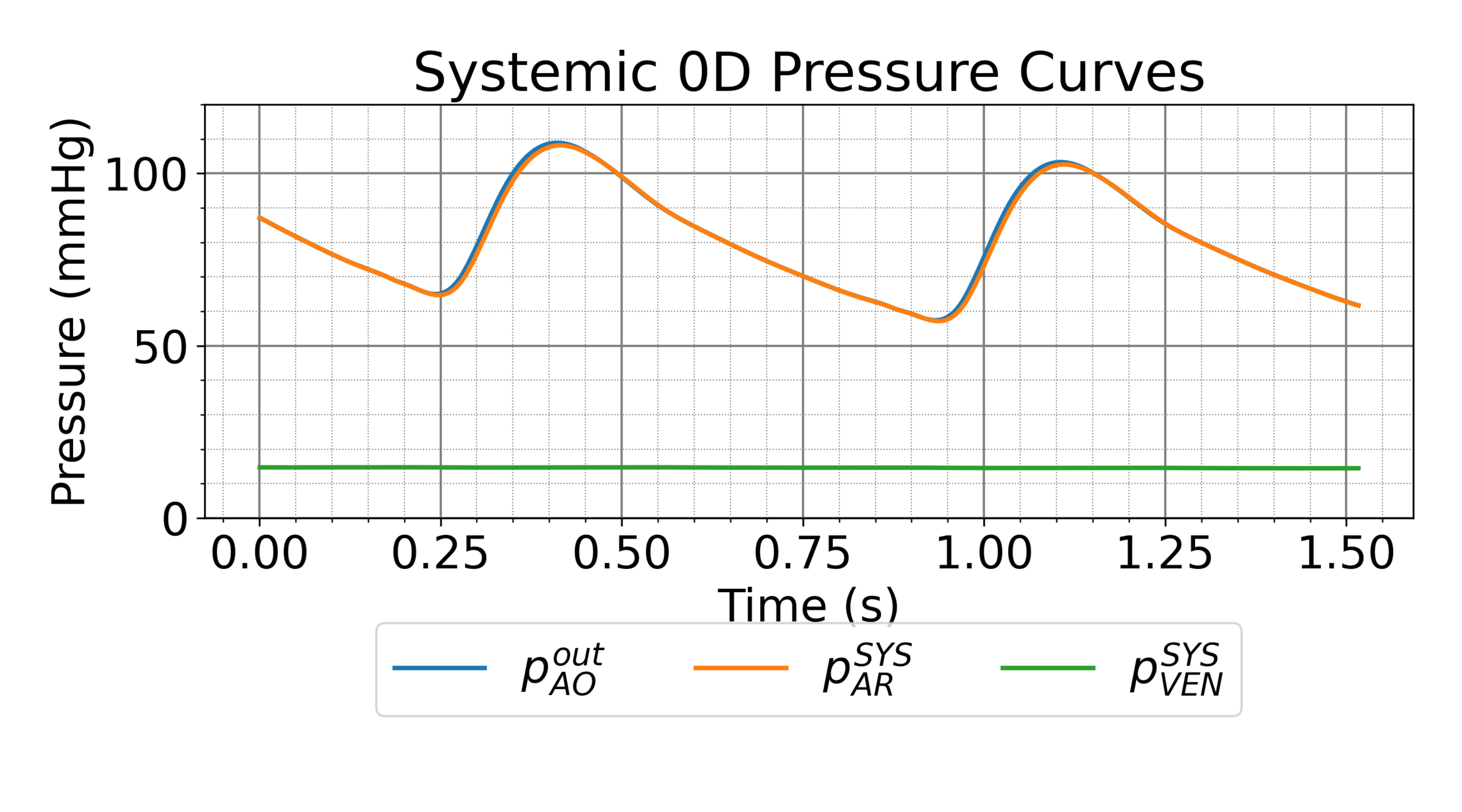}
  \end{subfigure}
  \hfill
  \begin{subfigure}[t]{0.49\textwidth}
    \centering
    \includegraphics[width=\linewidth]{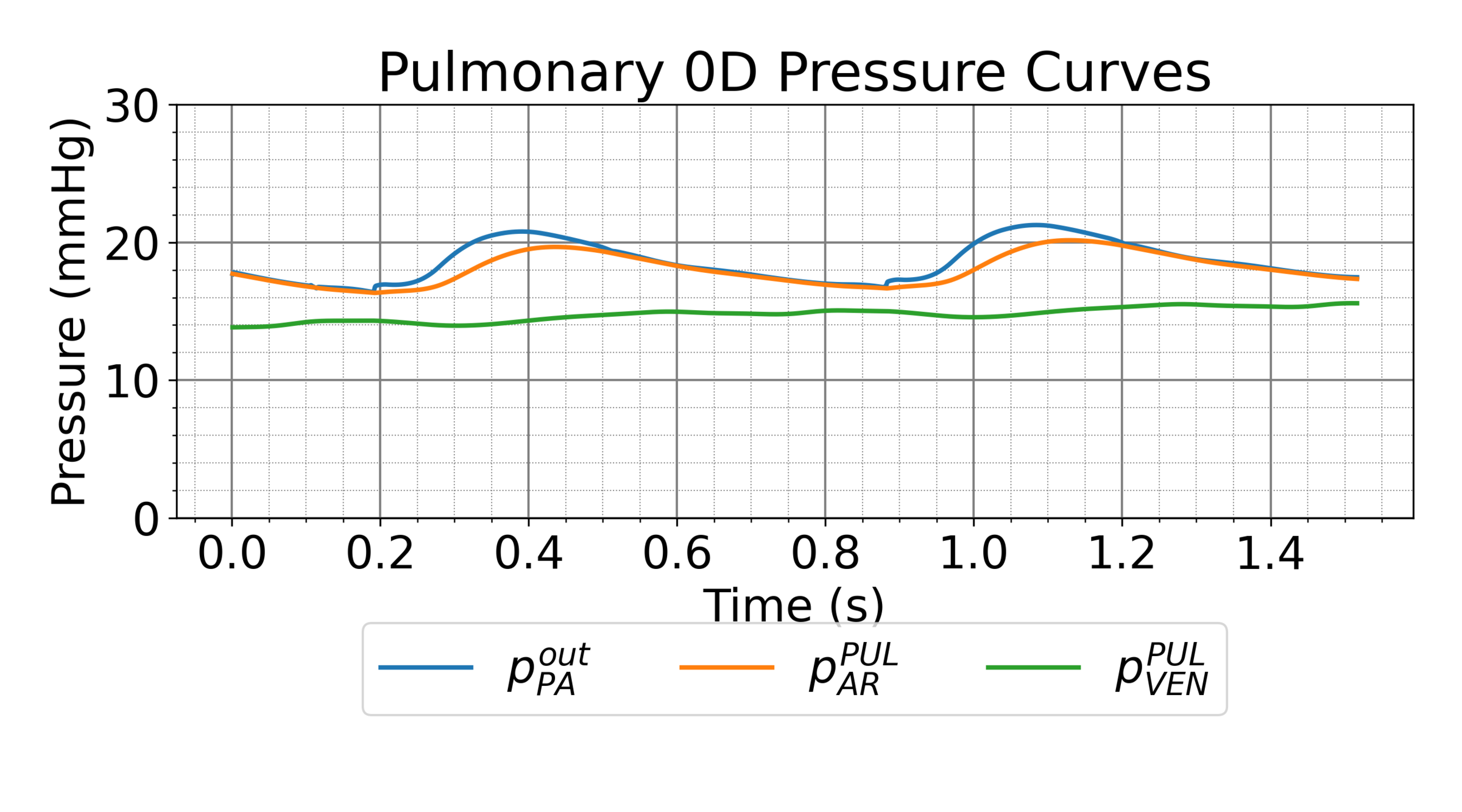}
  \end{subfigure}
  \caption{Systemic and pulmonary arterial pressure curves computed by the closed-loop 0D LPN}
  \label{fig:healthy:0d}
\end{figure}

\begin{figure}[h!]
    \centering
    \includegraphics[width=\linewidth]{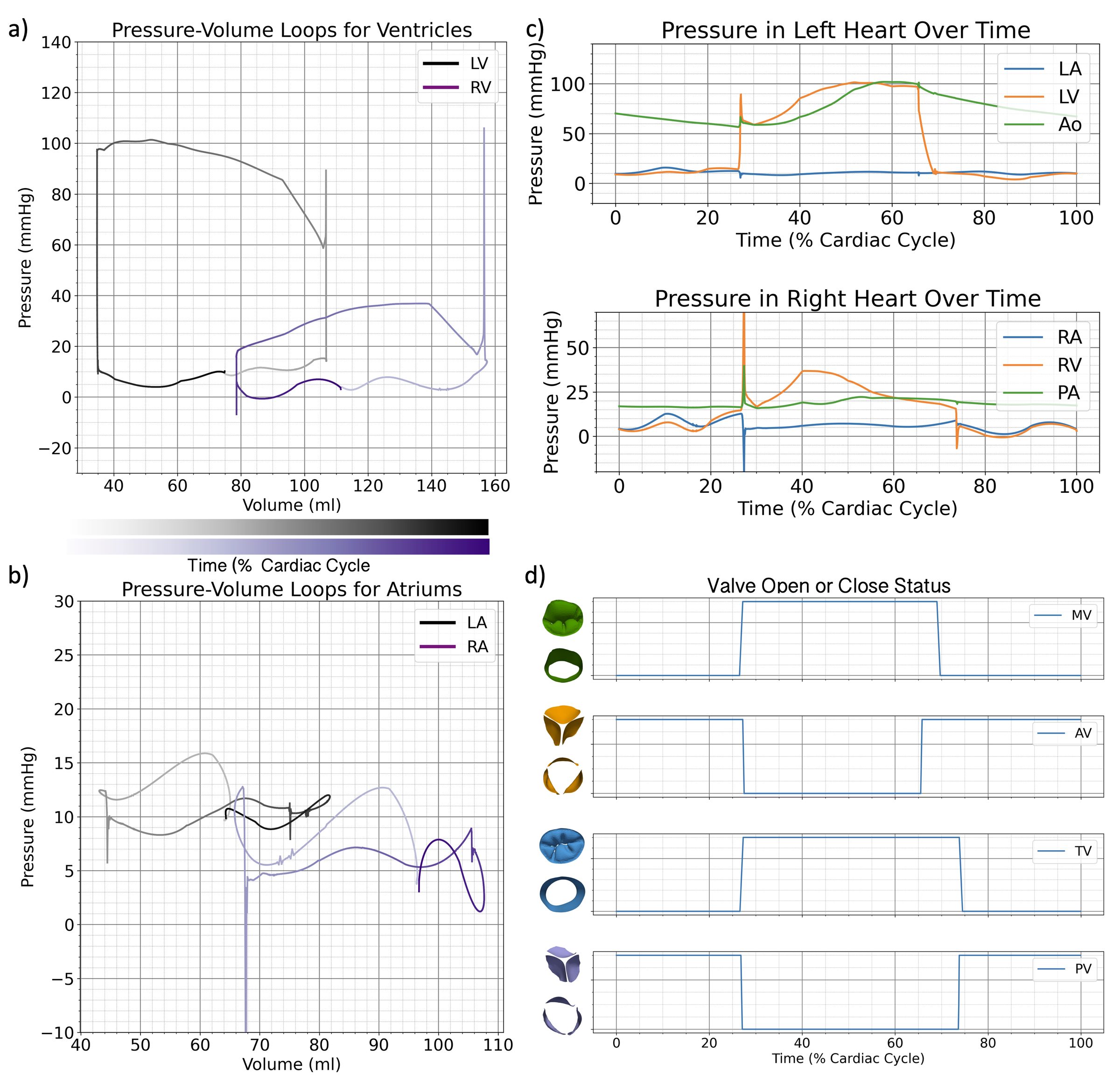}
    \caption{Simulation results for the healthy subject. (a) Pressure–volume loops for the left and right ventricles. (b) Pressure–volume loops for the left and right atria. The change in opacity indicates the time progression within the cardiac cycle. (c) Mean pressure curves for all chambers and great vessels over time. (d) Corresponding valve status showing the opening and closing phases over the same time.}
    \label{fig:healthy_curves}
\end{figure}

Figure~\ref{fig:healthy:0d} shows the systemic and pulmonary arterial pressure waveforms over multiple cardiac cycles computed by the closed-loop 0D LPN. Using the above LPN parameters and initial conditions, the 0D model reached a periodic solution with nearly identical pressure waveforms across successive cardiac cycles. Figure \ref{fig:healthy_curves} shows the pressure curves and the pressure-volume relationships extracted from the 3D simulation results during the last simulated cardiac cycle. Those pressure values are the mean pressure over the corresponding chamber or great vessel. The ventricular PV loops display standard rectangular or trapezoidal shapes. The atrial PV loops display both the a-loop, associated with atrial contraction, and the V loop, associated with passive filling. Figure \ref{fig:healthy_curves}d also shows the valve status changes between open and closed configurations triggered by the pressure or flow rate-based checking criteria of the RIS. Table \ref{tab:healthy_valve} displays the timing of the valve status change during the cardiac cycle. The timing of valve opening and closure followed the expected physiological sequence. At the onset of systole, the atrioventricular valves closed first, followed by the opening of the semilunar valves to permit ventricular ejection. At the onset of diastole, the semilunar valves closed first, followed by the opening of the atrioventricular valves to allow ventricular filling.

\begin{table}[h!]
\centering
\caption{Valve opening and closing timings extracted from simulations (\% of cardiac cycle) for the healthy case.}
\label{tab:healthy_valve}
\begin{tabular}{lcccc}
\hline
              & MV     & AV     & TV     & PV     \\ \hline
Open to close & 26.8\% & 65.6\% & 26.9\% & 74.1\% \\
Close to open & 69.3\% & 27.3\% & 73.7\% & 26.9\% \\ \hline
\end{tabular}
\end{table}

\begin{figure}[h!]
    \centering
    \begin{subfigure}[b]{\textwidth}
        \centering
        \includegraphics[width=\textwidth]{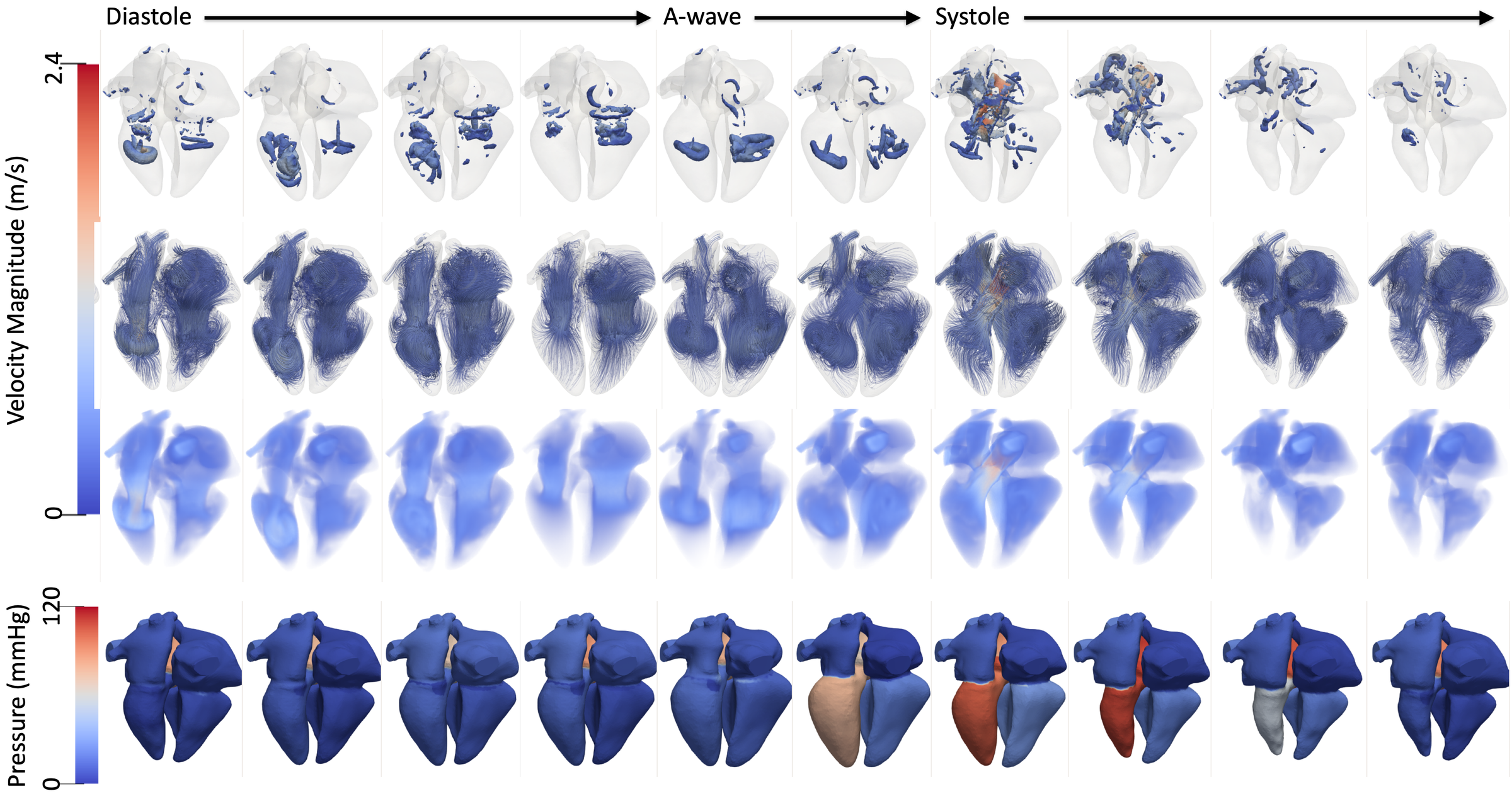}
        \caption{Anterior view }
        \label{fig:healthy_volume_front}
    \end{subfigure}
    \hfill 
    \begin{subfigure}[b]{\textwidth}
        \centering \includegraphics[width=\textwidth]{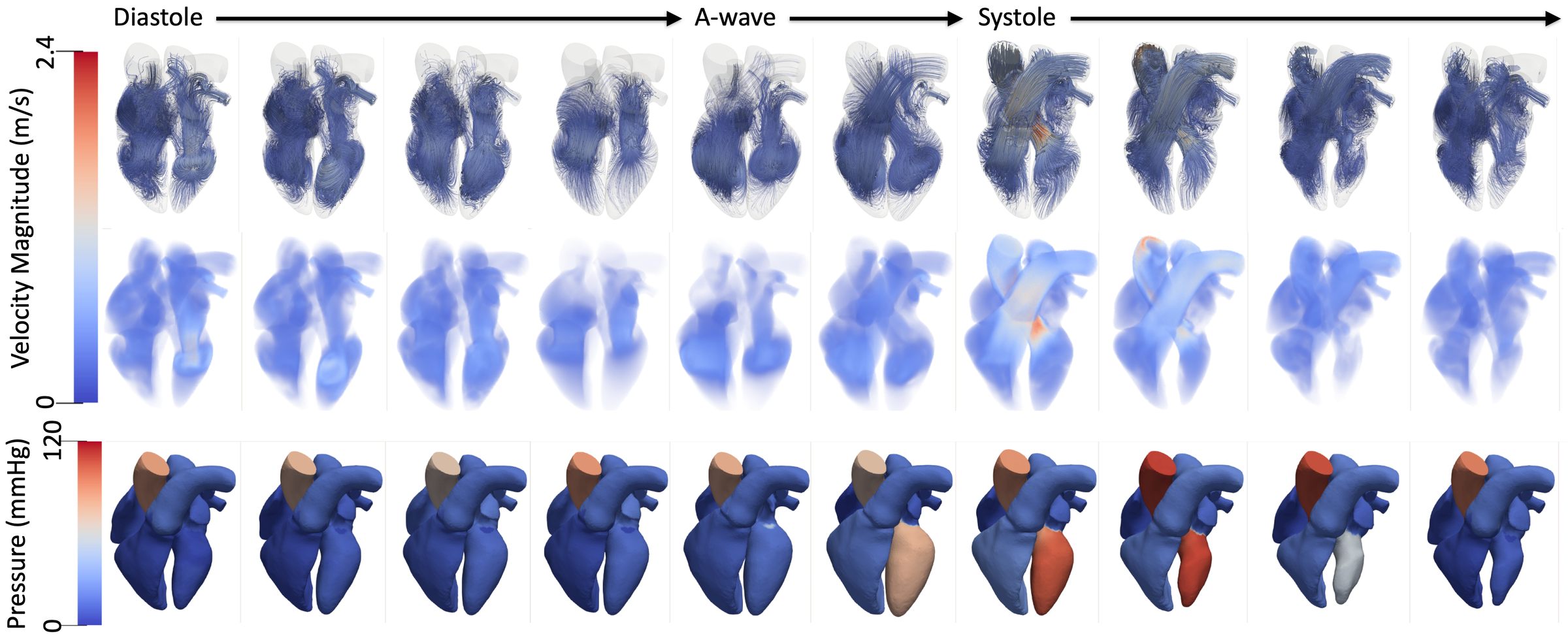}
        \caption{Posterior view }
        \label{fig:healthy_volume_back}
    \end{subfigure}
    
    \caption{Visualization of simulated velocity and pressure fields for the healthy subject. From top to bottom: Q-criterion iso-surface (threshold = $4000~\mathrm{s}^{-2}$), velocity streamlines, volumetric rendering of velocity magnitude, and pressure distribution on the surface.}
\label{fig:healthy_volume_plots}
\end{figure}

Although the simulation successfully reproduced physiologic cardiac behavior, we observed a significant pressure overshoot in both ventricles during the isovolumetric contraction phase, followed by a mild pressure undershoot and oscillation during the isovolumetric relaxation phase. In addition, the isovolumetric phases, which are typically defined by the simultaneous closure of the atrioventricular and semilunar valves, were extremely short in our simulations. The atrial pressure curves exhibited kinks during atrial contraction, and the PV-loop segments corresponding to the filling phases of both ventricles and atria showed pressure fluctuations rather than the expected steady pressure increase.  These artifacts are likely due to inaccuracies in prescribed cardiac motion, which was generated using cubic-spline interpolation of image-derived displacements with limited temporal resolution. Such pressure artifacts are particularly present during the iso-volumetric phases, since our approach does not explicitly enforce constant ventricular volume, and small spurious volume changes may occur.

Figure \ref{fig:healthy_volume_plots} visualizes the simulated velocity and pressure fields during a cardiac cycle for the healthy subject. The first and second rows show the instantaneous velocity structures, including Q-criterion iso-surfaces highlighting regions of vortex formation and streamlines visualizing the instantaneous blood flow trajectories. When ventricular pressure falls below atrial pressure, the resulting pressure gradient causes the mitral and tricuspid valves to open, allowing blood to flow from the atria into the left and right ventricles. During early diastole, inflow through the mitral and tricuspid valves generates strong vortical structures within both ventricles, characterized by the formation of vortex rings surrounding the inflow jets. This early filling phase is largely passive and is driven by ventricular relaxation and the associated pressure gradient. As passive inflow continues, atrial and ventricular pressures gradually equalize, leading to a reduction in forward flow and the breakdown of the initially coherent vortex rings into smaller, less organized flow structures. This period of reduced transvalvular flow defines diastasis and persists until the onset of atrial contraction. Subsequent atrial contraction provides an additional filling, generating secondary vortex rings around the inflow jets that propagate toward the ventricular apex.

\begin{figure}[h!]
    \centering
    \includegraphics[width=\textwidth]{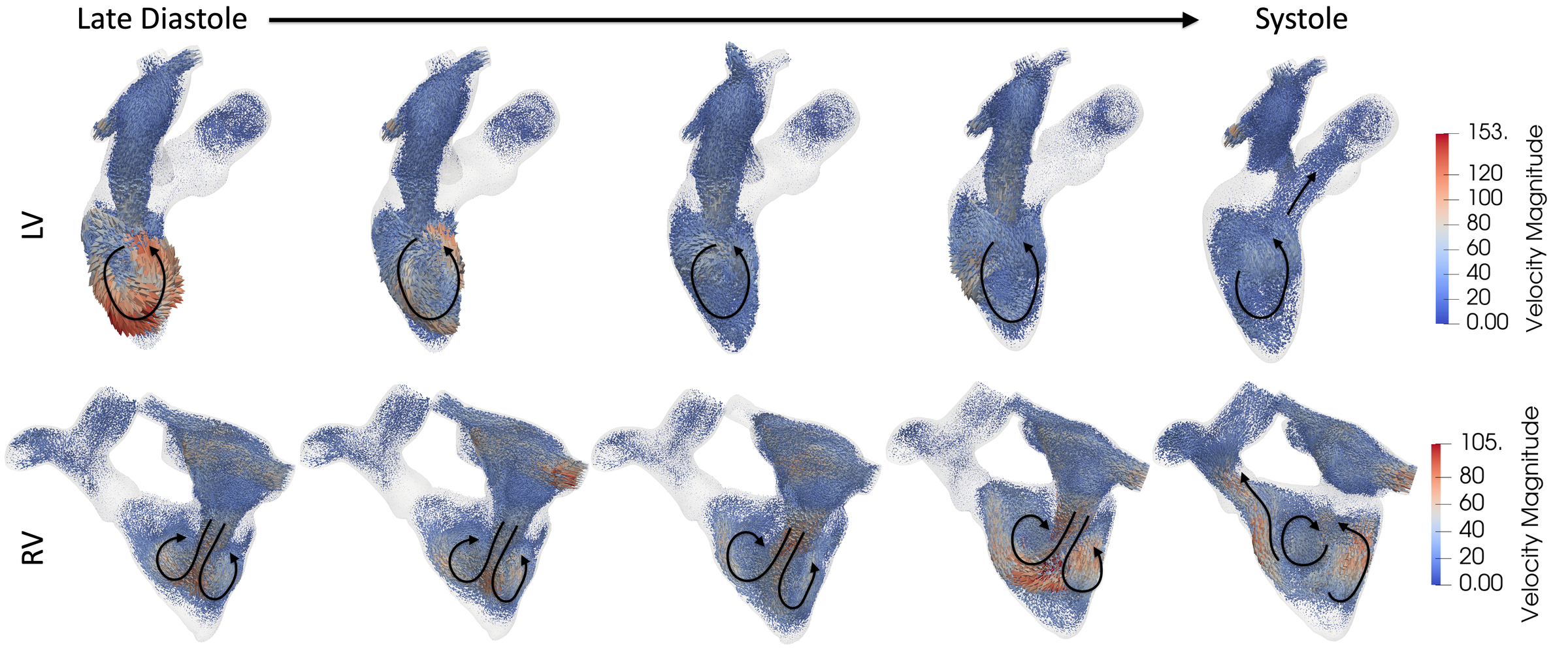}
    \caption{Flow patterns in the LV, LA, aorta (top) and in the RV, RA, PA (bottom) for the healthy subject during late diastole. The color map indicates velocity magnitude (cm/s) }
    \label{fig:diastole_pattern_healthy}
\end{figure}

Figure~\ref{fig:diastole_pattern_healthy} further shows the intraventricular flow patterns during atrial contraction (a-wave) at late diastole and the transition into systole in the healthy subject. Atrial contraction produces an additional inflow into both ventricles, forming a well-defined inflow jet and organized vortical structures. In the LV, the a-wave inflow is associated with a dominant counterclockwise circulation. This structure persists through late diastole and redirects flow from the inflow jet toward the posterior wall, through the apical region, and then toward the outflow tract. In the RV, late-diastolic filling is characterized by more complex flow patterns due to the difference in geometry between RV and LV. The a-wave inflow generates two dominate vortical structures, with flow recirculation developing along the RV free wall and toward the outflow region. As the cardiac cycle progresses toward systole, the late-diastolic vortices in both ventricles weaken and reorganize, with flow gradually redirected toward the outflow tracts in preparation for ventricular ejection.

During systole ejection, the ventricular flow exhibits a converging pattern with a rapid increase in velocity as the blood is directed through the aortic and the pulmonary outflow tracts. Vortical structures are also observed in the aorta and PA during the systolic contraction. The lower panels of Figure \ref{fig:healthy_volume_plots} show the pressure distribution on the endocardial surfaces. Ventricular pressures rise sharply during systole, reaching approximately 110 mmHg in the LV and 25 mmHg in the RV, consistent with physiologic values. Overall, the RIS–ALE simulation captures realistic spatiotemporal evolution of intracardiac velocity and pressure fields, reproducing physiologic phasing of the cardiac flow cycle.


\subsection{Simulation results for the CHD patient }

\begin{figure}[h!]
  \centering
  \begin{subfigure}[t]{0.49\textwidth}
    \centering
    \includegraphics[width=\linewidth]{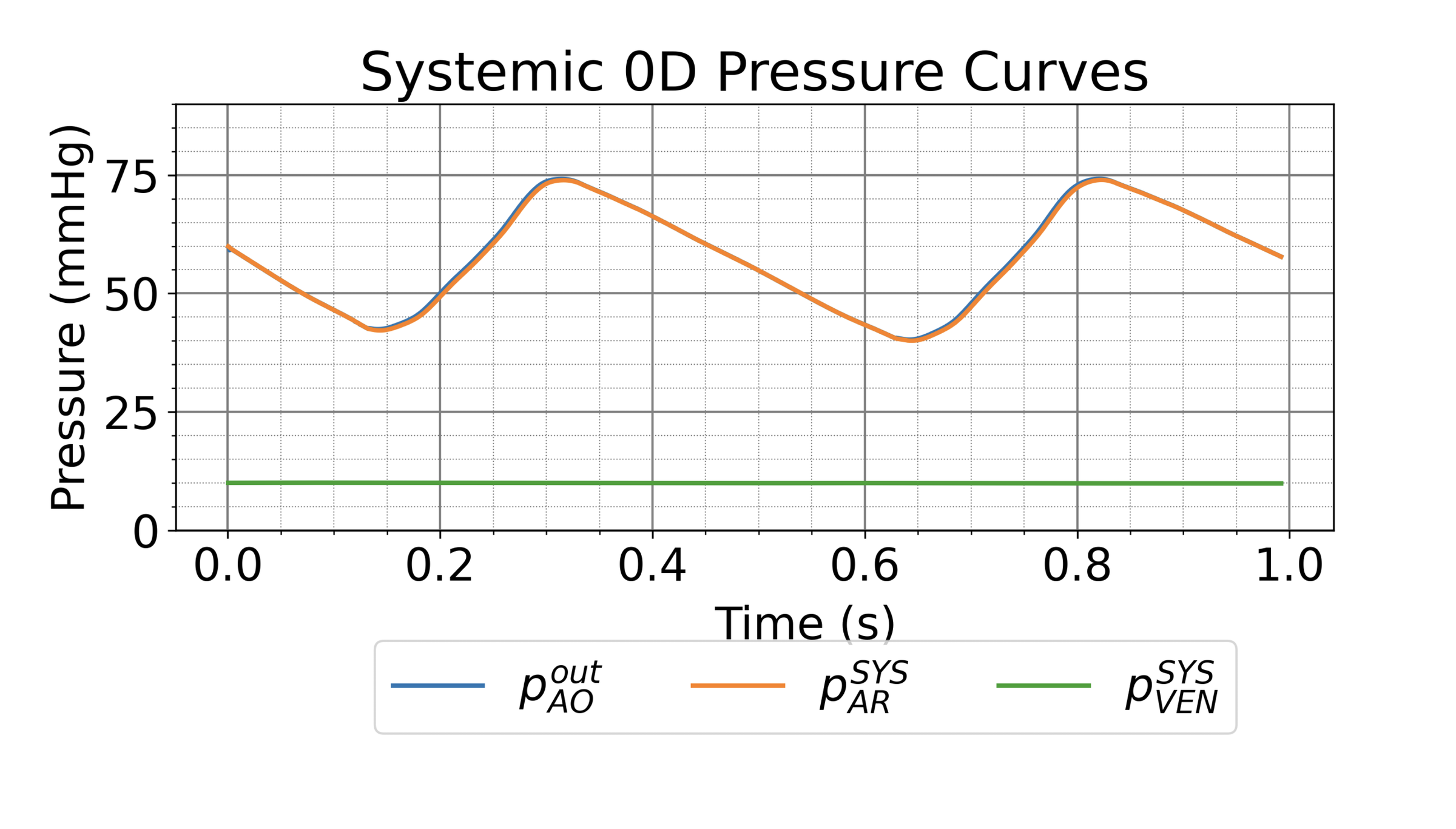}
  \end{subfigure}
  \hfill
  \begin{subfigure}[t]{0.49\textwidth}
    \centering
    \includegraphics[width=\linewidth]{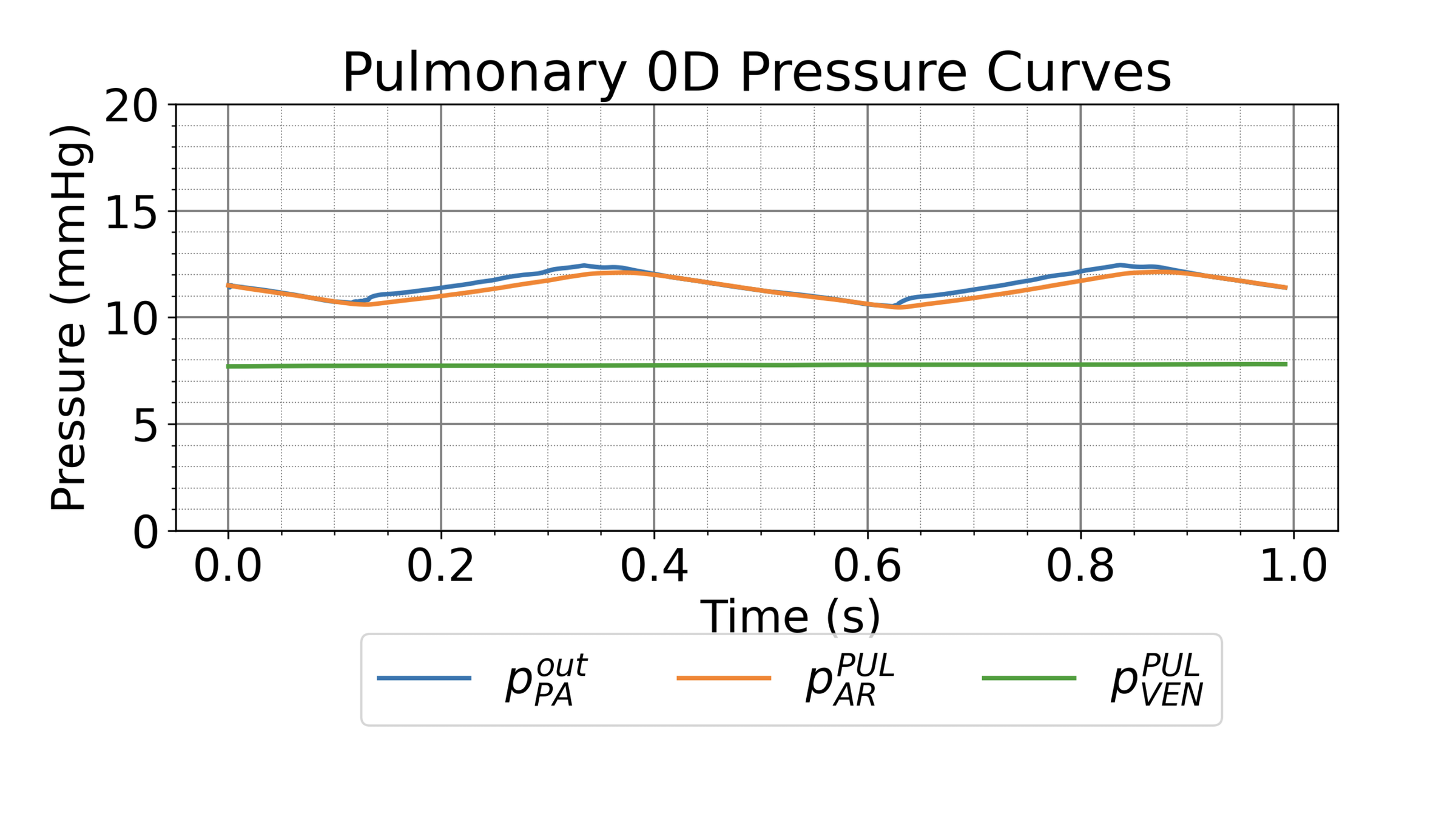}
  \end{subfigure}
  \caption{Systemic and pulmonary arterial pressure curves computed by the closed-loop 0D LPN for the CHD patient.}
  \label{fig:vsd:0d}
\end{figure}

Table \ref{tab:0D_params_chd} summarizes the parameters used in the closed-loop 0D LPN for the CHD case. These parameters were adjusted from Brown \emph{et al.} \cite{brown2025biventricular} so that the simulated pressure matched the patient-specific pressure measurements obtained from the cardiac catheterization procedure. Figure~\ref{fig:vsd:0d} shows the systemic and pulmonary arterial pressure waveforms over multiple cardiac cycles computed by the closed-loop 0D LPN for this CHD patient. Using the above LPN parameters and initial conditions, the 0D model reached a periodic solution with nearly identical pressure waveforms across successive cardiac cycles.

\renewcommand{\arraystretch}{1.2} 
\begin{longtable}{l c c c}
\caption{Parameters of the 0D circulation model for the CHD case. } \label{tab:0D_params_chd} \\
\hline
\textbf{Parameter} & \textbf{Symbol} & \textbf{Value} & \textbf{Units} \\
\hline
\endfirsthead

\multicolumn{4}{c}%
{\tablename\ \thetable\ -- \textit{continued from previous page}} \\
\hline
\textbf{Parameter} & \textbf{Symbol} & \textbf{Value} & \textbf{Units} \\
\hline
\endhead

\hline \multicolumn{4}{r}{\textit{Continued on next page}} \\
\endfoot

\hline
\endlastfoot
\textbf{\textit{Systemic circulation}}  \\
Systemic arterial resistance & $R_\mathrm{AR}^\mathrm{SYS}$ &  $1.578 $ & mmHg s mL$^{-1}$ \\
Systemic arterial capacitance & $C_\mathrm{AR}^\mathrm{SYS}$ & $0.290 $ & mL mmHg$^{-1}$ \\
Systemic arterial inductance & $L_\mathrm{AR}^\mathrm{SYS}$ & $0.6$ & mmHg s$^2$ mL$^{-1}$ \\
Systemic venous resistance & $R_\mathrm{VEN}^\mathrm{SYS}$ & $0.315 $ & mmHg s mL$^{-1}$ \\
Systemic venous capacitance & $C_\mathrm{VEN}^\mathrm{SYS}$ & $120.0$ & mL mmHg$^{-1}$ \\
Systemic venous inductance & $\displaystyle L_\mathrm{VEN}^\mathrm{SYS}$ & $5\times10^{-5}$ & mmHg s$^2$ mL$^{-1}$ \\

\\
\textbf{\textit{Pulmonary circulation}}  \\
Pulmonary arterial resistance & $R_\mathrm{AR}^\mathrm{PUL}$ & $0.136$ & mmHg s mL$^{-1}$ \\
Pulmonary arterial capacitance & $C_\mathrm{AR}^\mathrm{PUL}$ & $4.0$ & mL mmHg$^{-1}$ \\
Pulmonary arterial inductance & $L_\mathrm{AR}^\mathrm{PUL}$ & $0.02$ & mmHg s$^2$ mL$^{-1}$ \\
Pulmonary venous resistance & $R_\mathrm{VEN}^\mathrm{PUL}$ & $0.05$ & mmHg s mL$^{-1}$ \\
Pulmonary venous capacitance & $C_\mathrm{VEN}^\mathrm{PUL}$ & $160.0$ & mL mmHg$^{-1}$ \\
Pulmonary venous inductance & $L_\mathrm{VEN}^\mathrm{PUL}$ & $1.25\times10^{-5}$ & mmHg s$^2$ mL$^{-1}$ \\

\\
\textbf{\textit{Initial conditions}}  \\
Systemic arterial pressure & $p_\mathrm{AR}^\mathrm{SYS}(0)$ & $60.00$ & mmHg \\
Systemic venous pressure & $p_\mathrm{VEN}^\mathrm{SYS}(0)$ & $7.70 $ & mmHg \\
Pulmonary arterial pressure & $p_\mathrm{AR}^\mathrm{PUL}(0)$ & $11.50$ & mmHg \\
Pulmonary venous pressure & $p_\mathrm{VEN}^\mathrm{PUL}(0)$ & $10.00$ & mmHg \\
Systemic arterial flow & $Q_\mathrm{AR}^\mathrm{SYS}(0)$ & $40.00$ & mL s$^{-1}$ \\
Pulmonary arterial flow & $Q_\mathrm{AR}^\mathrm{PUL}(0)$ & $35.00$ & mL s$^{-1}$ \\
\hline
\end{longtable}

\begin{figure}[h]
    \centering
    \includegraphics[width=\linewidth]{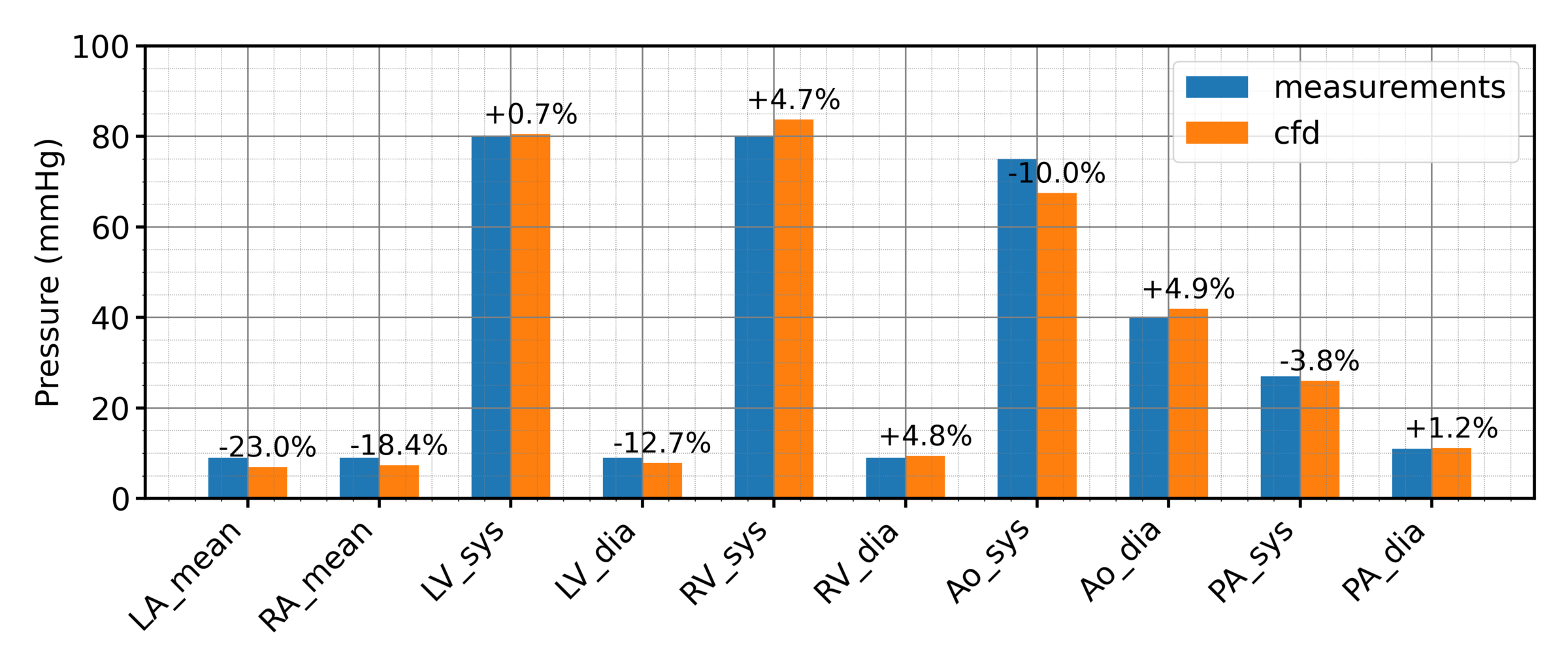}
    \caption{Comparison of simulated chamber pressures with cardiac catheterization data. For consistency with the catheter measurements, CFD pressures for the LA and RA were computed as mean values over the full cardiac cycle, while LV, RV (functional), PA, and aortic pressures were averaged during systole or diastole.}
    \label{fig:cfd_cath_compare}
\end{figure}

Figure~\ref{fig:cfd_cath_compare} demonstrates good agreement between the simulated chamber pressures and the catheterization measurements. Specifically, peak systolic and late-diastolic pressures were compared for the functional LV, functional RV, aorta, and PA, while mean pressures were compared for the LA and RA. For most chambers, the pressure differences between simulation and catheterization data were within 10\%. Larger discrepancies were observed for the mean LA pressure (23.0\%), mean RA pressure (18.4\%), and the late-diastolic pressure of the functional LV (morphological RV), which differed by 12.7\%. The simulation reproduced the elevated and nearly equal systolic pressures in both ventricles ($\approx 80 \text{mmHg}$) and the large pressure gradient across the pulmonary valve, consistent with pulmonary stenosis. The systemic arterial pressure in this pediatric patient was moderately lower than in the healthy adult case, while the systemic pulmonary arterial pressure was substantially elevated. 

\begin{figure}[h!]
    \centering
    \includegraphics[width=\linewidth]{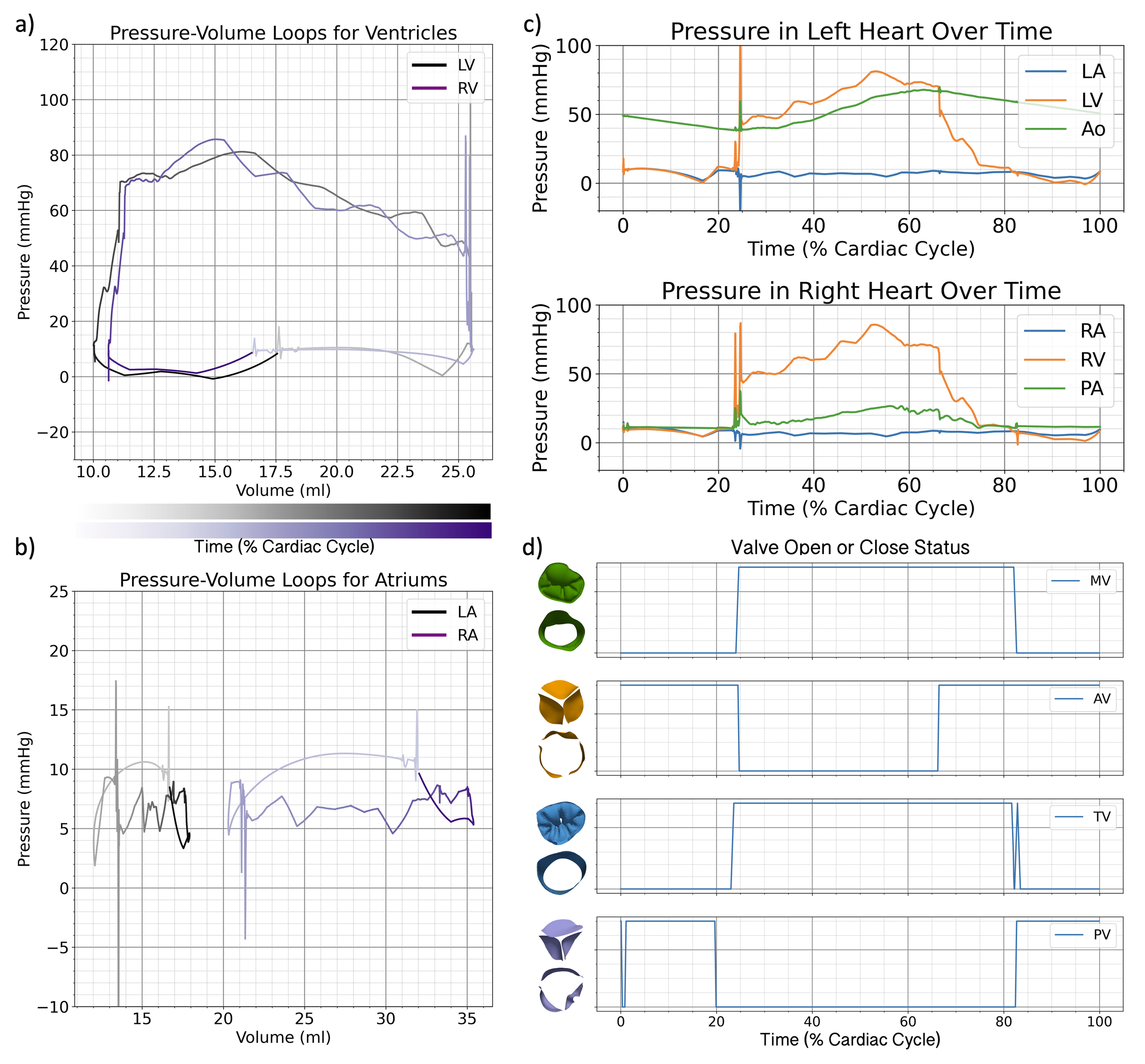}
    \caption{Simulation results for the CHD patient. (a) Pressure–volume loops for the left and right ventricles. (b) Pressure–volume loops for the left and right atria. The change in opacity indicates the time progression within the cardiac cycle. (c) Mean pressure curves for all chambers and great vessels over time. (d) Corresponding valve status showing the opening and closing phases over the same time period.}
    \label{fig:vsd_curves}
\end{figure}

Figure \ref{fig:vsd_curves} summarizes the simulated cardiac pressures, PV-loops, and valve open and closing status for the CHD patient. The RV here refers to the functional RV (morphological LV), and the LV refers to the functional LV (morphological RV). The RV and LV PV-loops are nearly identical, demonstrating substantially elevated RV pressure compared with the healthy case. This also indicates that the pressure gradient across the VSD is roughly small and consistent throughout the cardiac cycle. The atrial PV loops retain identifiable a-loops (atrial contraction) and v-loops (atrial filling), with the pressure value between the two atria very similar except for a much larger volume in the RA than in the LA. The time-resolved pressure curves (Figure \ref{fig:vsd_curves}c) demonstrate a large pressure gradient between the RV and the pulmonary artery during systole, which is consistent with the narrowed PA seen in the model compared with the healthy cases. The valve status plots (Figure \ref{fig:vsd_curves}d) show, in general, physiologic sequencing of opening and closure similar to what was seen in the healthy cases. However, our simulation successfully captured a longer systolic duration than diastole, which is expected given the elevated heart rate of 123 beats per minute, where diastole is naturally shortened. Table~\ref{tab:chd_valve} summarizes the timing of valve opening and closure over the cardiac cycle for this CHD patient. At the onset of diastole, the semilunar valves closed first, followed by the opening of the atrioventricular valves to permit ventricular filling. For this patient, the simulation captured a longer, and more physiologically realistic, interval during which both the AV and MV remained closed. Specifically, the interval between AV close and MV open (16.1\% of the cardiac cycle) was substantially longer than the corresponding interval between PV close and TV open (0.6\% of the cardiac cycle), and was also longer than that observed in the healthy case, where this phase is much shorter. At the onset of systole, the MV closed first, followed by opening of the AV to allow ventricular ejection. In the normal case, the TV closes before the PV opens during RV ejection. However, for this patient, the simulation predicted that the PV opened before TV closure, as RV pressure exceeded PA pressure prior to TV closure. 

Similar to the healthy case, we observed a significant pressure overshoot in both ventricles during the isovolumetric contraction phase, as well as extremely short isovolumetric phases. The pressure curves also exhibited kinks and oscillations, particularly during systole, which constitutes the majority of the cardiac cycle. Similarly, these artifacts are likely due to inaccuracies in the prescribed cardiac motion. In this CHD case, the issue is more pronounced, as evidenced by the presence of pressure oscillations that are not observed in the healthy case, because the 4D-flow MRI provides a higher temporal resolution (~30 time steps) than the CT data used in the healthy case (~10 time steps), thereby introducing greater temporal inconsistency in the prescribed motion.

\begin{table}[h!]
\centering
\caption{Valve opening and closing timings extracted from the CHD simulations (\% of cardiac cycle).}
\label{tab:chd_valve}
\begin{tabular}{lcccc}
\hline
              & MV     & AV     & TV     & PV     \\ \hline
Open to close & 24.3\% & 66.2\% & 23.2\% & 82.5\% \\
Close to open & 82.3\% & 24.6\% & 83.1\% & 19.7\% \\ \hline
\end{tabular}
\end{table}



\begin{figure}[h!]
    \centering
    \begin{subfigure}[b]{\textwidth}
        \centering
        \includegraphics[width=\textwidth]{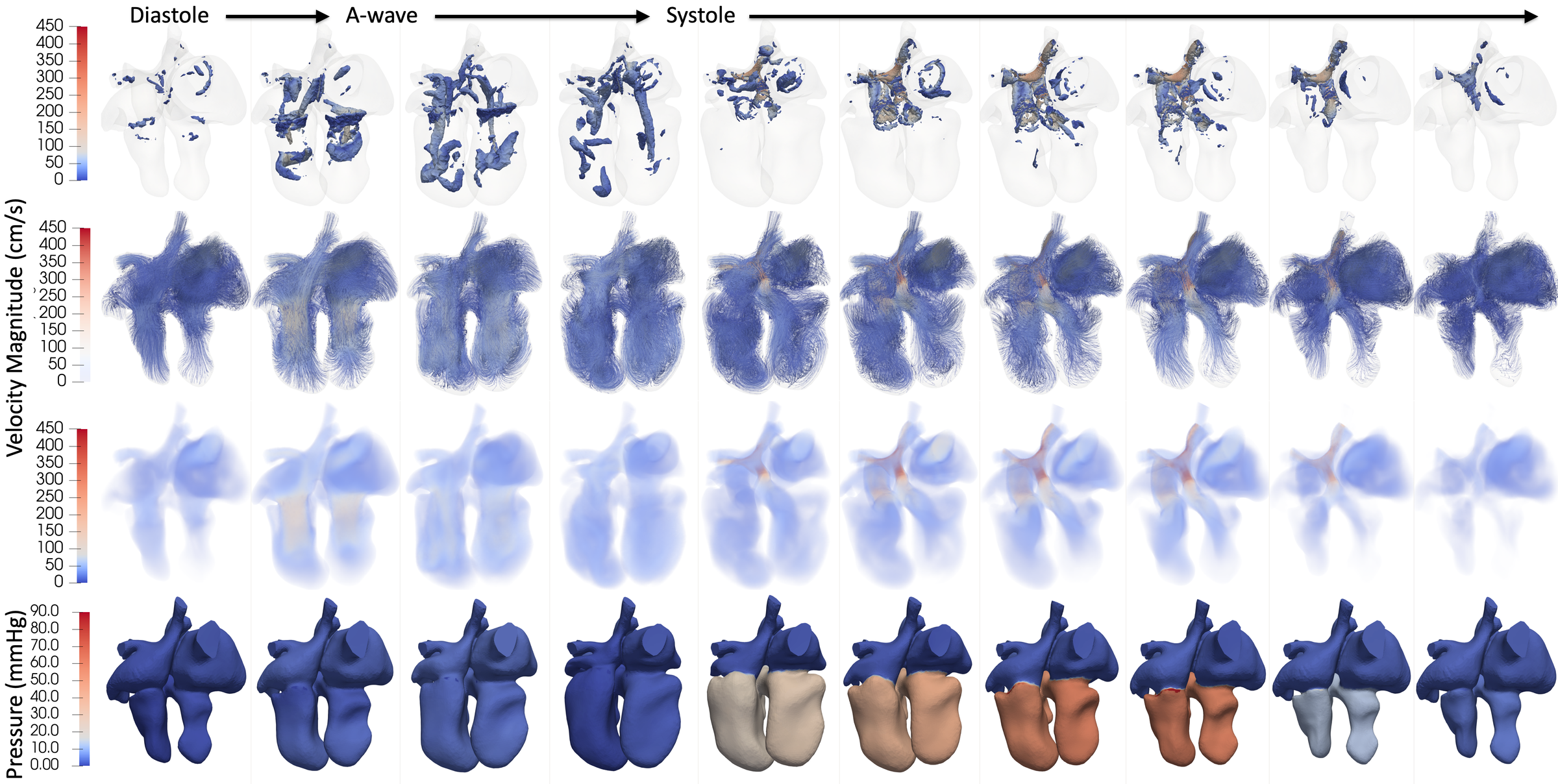}
        \caption{Anterior view }
        \label{fig:vsd_volume_front}
    \end{subfigure}
    \hfill 
    \begin{subfigure}[b]{\textwidth}
        \centering \includegraphics[width=\textwidth]{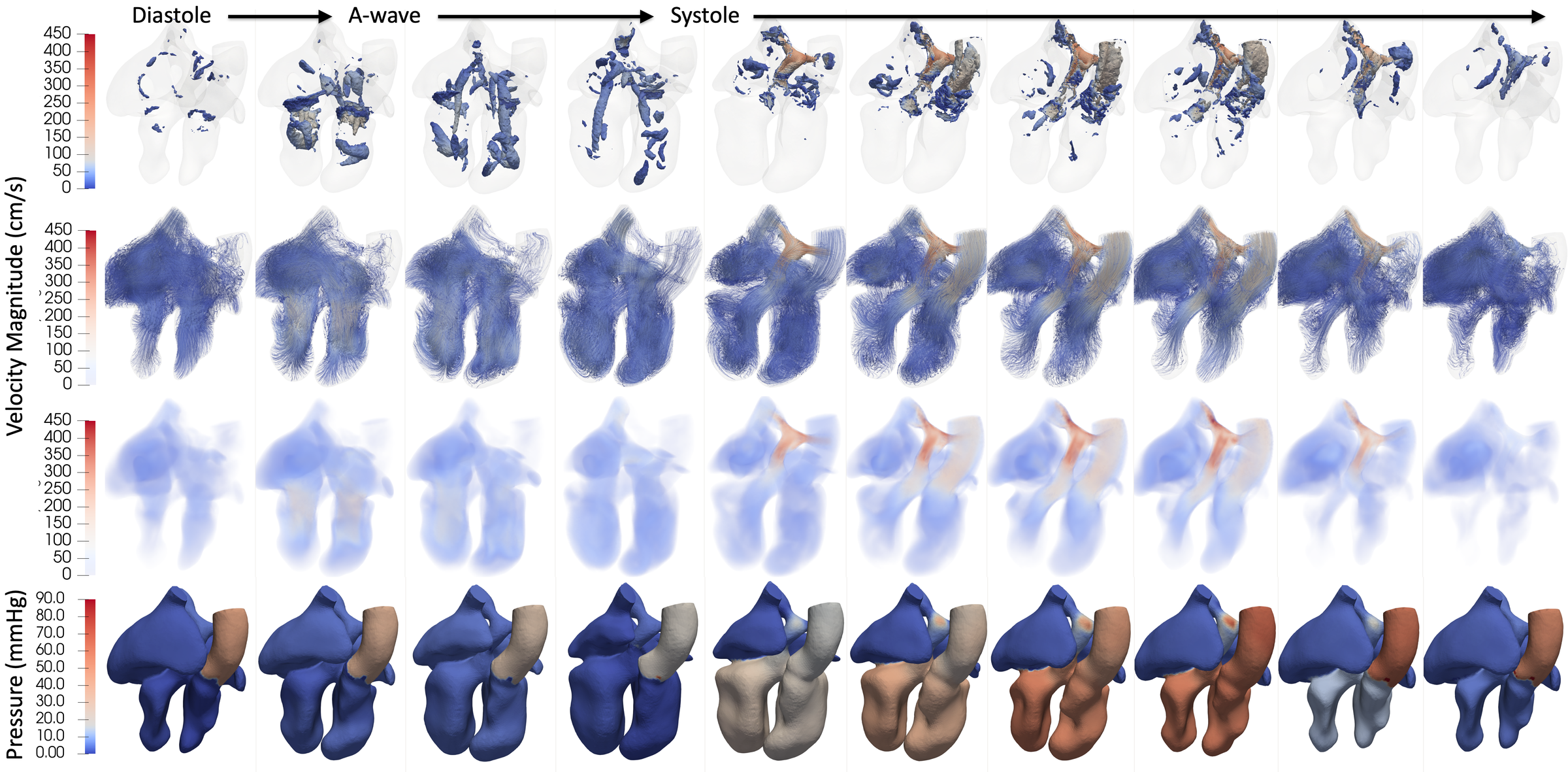}
        \caption{Posterior view }
        \label{fig:vsd_volume_back}
    \end{subfigure}
    
    \caption{Visualization of simulated velocity and pressure fields for the CHD patient. From top to bottom: Q-criterion iso-surface (threshold = $100~\mathrm{s}^{-2}$), velocity streamlines, volumetric rendering of velocity magnitude, and pressure distribution on the surface.}
\label{fig:vsd_volume_plots}
\end{figure}

Figure~\ref{fig:vsd_volume_plots} visualizes the simulated velocity and pressure fields over a cardiac cycle for the CHD patient. The first and second rows show instantaneous velocity structures, including Q-criterion isosurfaces highlighting regions of vortex formation and streamlines illustrating instantaneous blood flow trajectories. As in the healthy case, when ventricular pressure falls below atrial pressure, the resulting pressure gradient causes the MV and TV to open, allowing blood to flow from the atria into the left and right ventricles. Compared with the healthy case, however, this CHD patient exhibits a shorter diastolic filling phase and a less pronounced vortex ring surrounding the inflow jet. During systolic ejection, the ventricular flow shows a converging pattern similar to that observed in the healthy subject, with velocities increasing rapidly as blood is directed toward the aortic and pulmonary outflow tracts. For this CHD patient, the pulmonary artery has a smaller diameter than the aorta, together with elevated RV pressure, leading to a steeper pressure gradient between the ventricle and the vessel and resulting in a faster outflow jet in the PA than in the aorta. Furthermore, due to the presence of a large VSD, the pressures in the LV and RV remain comparable. As a result, unlike in some VSD cases where inter-ventricular shunting is prominent, this patient does not exhibit sustained or pronounced flow shunting between the ventricles over the cardiac cycle.

\begin{figure}[h!]
    \centering
    \includegraphics[width=\textwidth]{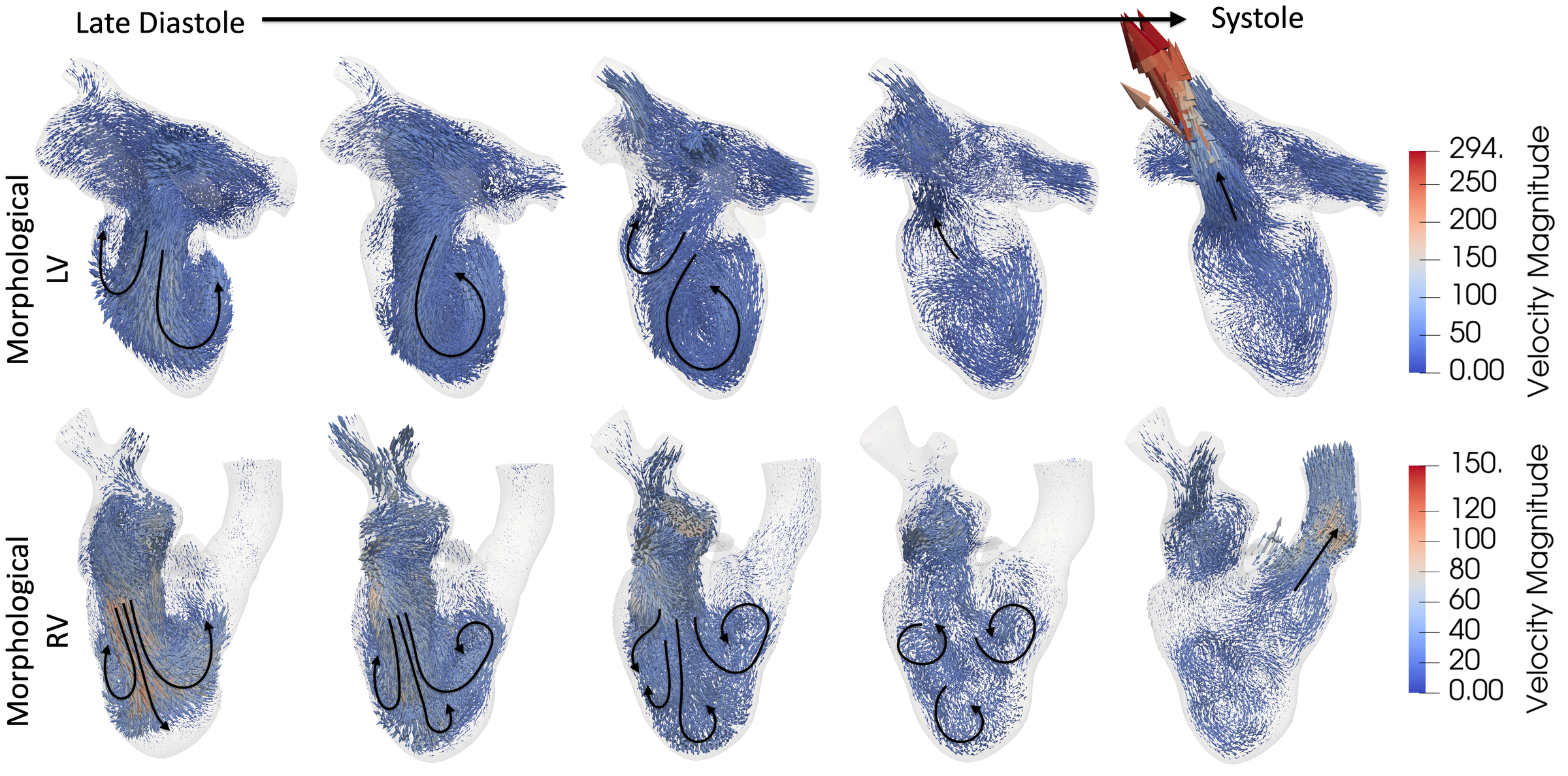}
    \caption{Flow patterns in the morphological LV, RA and PA (top) and in the morphological RV, LA, and aorta (bottom) for the CHD patient during late diastole. The color map indicate velocity magnitude (cm/s)}
    \label{fig:diastole_pattern_vsd}
\end{figure}

Figure~\ref{fig:diastole_pattern_vsd} further illustrates the intraventricular flow patterns during late diastole (a-wave) and the transition into systole for the CHD patient. During late diastole, atrial contraction in both the RA and LA generates additional inflow jets that contribute to ventricular filling. In this patient, however, the RA fills into the morphological LV, while the LA fills into the morphological RV. The abnormal cardiac anatomy results in diastolic flow patterns that differ markedly from those observed in the healthy case. In the morphological LV (functional RV), late-diastolic inflow gives rise to a dominant recirculating flow structure. Unlike the clockwise vortex observed in the healthy LV, the circulation in the morphological LV is counterclockwise, with inflow first impinging on the anterior wall near the PA outlet, traveling through the apical region, and then redirecting back toward the PA. This vortical structure weakens toward the end of late diastole as the flow transitions toward systolic ejection through the PA. In contrast, the morphological RV connected to the LA exhibits more complex diastolic flow patterns characterized by multiple vortical structures. As the inflow jet enters the chamber, shear interactions with the surrounding flow slow the lateral portions of the jet, leading to roll-up and the formation of two dominant vortices along the ventricular walls. Meanwhile, the central portion of the inflow jet penetrates toward the apex, resulting in the presence of three major recirculating regions by the end of diastole. These flow structures persist briefly before reorganizing as the flow exits through the aorta during systole.

\begin{figure}[h!]
    \centering
    \begin{subfigure}[b]{\textwidth}
        \centering
        \includegraphics[width=\textwidth]{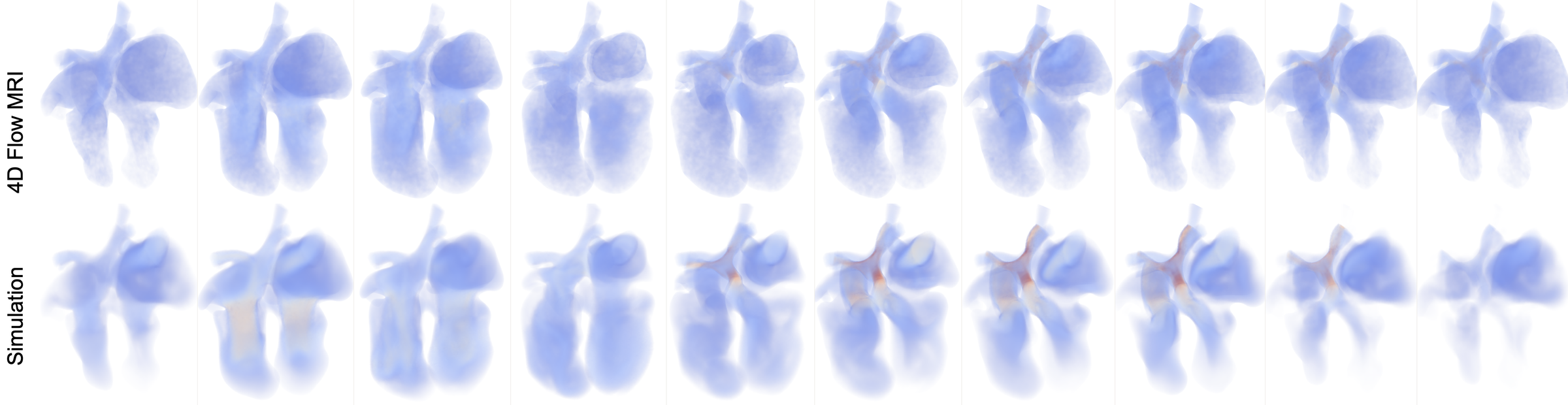}
        \caption{Anterior view }
        \label{fig:4dflow_volume_front}
    \end{subfigure}
    \hfill 
    \begin{subfigure}[b]{\textwidth}
        \centering \includegraphics[width=\textwidth]{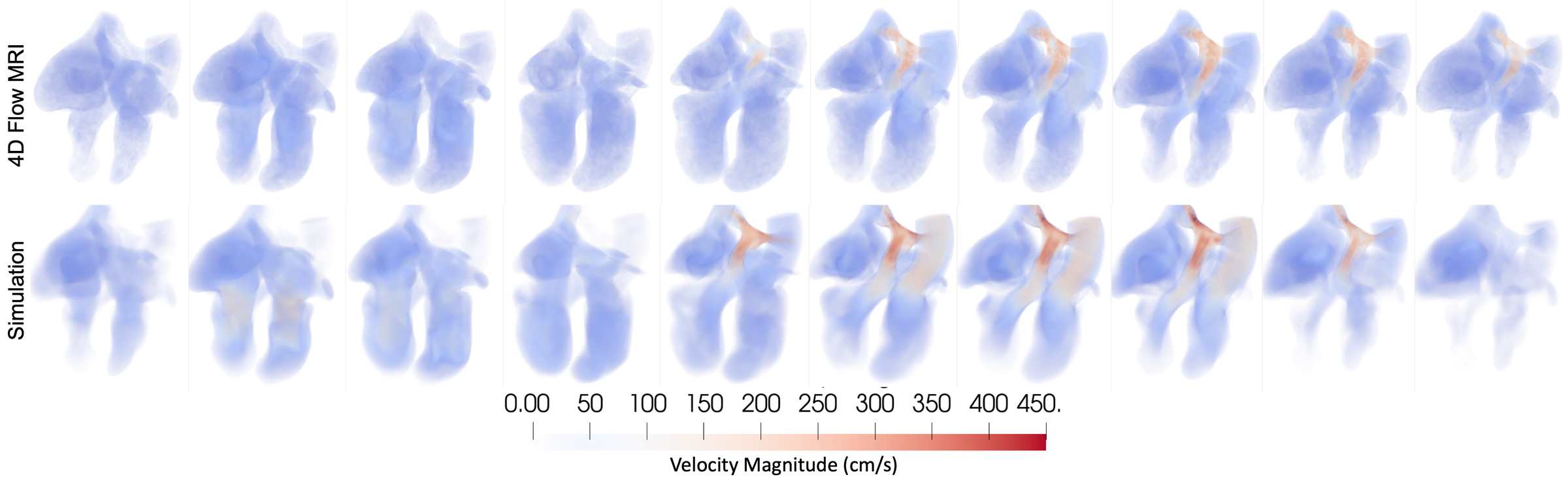}
        \caption{Posterior view }
        \label{fig:4dflow_volume_back}
    \end{subfigure}
    
    \caption{Comparison of simulated velocity fields with 4D-Flow MRI, shown as volumetric renderings at different phases of the cardiac cycle.}
\label{fig:4dflow_volume}
\end{figure}

Figure \ref{fig:4dflow_volume} compares the volumetric renderings of the velocity fields from the RIS–ALE simulation with 4D-Flow MRI projected onto the same mesh for the CHD patient. Across all cardiac phases, the simulation reproduced the overall spatial and temporal flow patterns observed in the 4D-Flow MRI data. Namely, the simulation reproduced diastolic inflow into the ventricles, as well as systolic ejection into the aorta and PA, occurring at similar timings within the cardiac cycle. The simulation also successfully reproduced the high outflow velocity into PA. However, the RIS-ALE simulation produced higher velocity magnitudes compared with 4D-Flow MRI during both diastolic inflow and systolic ejection. Furthermore, the pulmonary artery jet also persisted for a longer duration in the MRI than in the simulation. To provide an additional point of comparison, we further compared both CFD- and 4D Flow MRI-derived velocities against available echocardiographic measurements. Namely, using localized velocity extraction at anatomically matched sites, we compared the 99th-percentile (p99) velocity from the CFD field and the 4D flow MRI-derived velocity field against the available echocardiographic peak velocity measurements at the left PA, right PA, and TV (Table \ref{tab:echo}). For the branch PAs, sphere-based sampling yielded CFD p99 values of 3.717 m/s in the LPA and 4.271 m/s in the RPA, compared with 4D flow MRI values of 2.297 m/s and 2.772 m/s, and echo peak velocities of 4.2 m/s and 4.1 m/s, respectively. The CFD results were closer to the echocardiographic measurements in the branch pulmonary arteries, whereas 4D Flow MRI yielded lower peak velocities. At the TV inflow, the CFD p99 velocity was 1.125 m/s, modestly higher than the echocardiographic E-wave peak velocity of 0.9 m/s, while the 4D Flow MRI p99 velocity was lower at 0.623 m/s.

\begin{table}[ht]
\centering
\begin{tabular}{lccc}
\hline
Region & CFD  (m/s) & 4D Flow MRI  (m/s) & Echo (m/s) \\
\hline
LPA & 3.717 & 2.297 & 4.2 \\
RPA & 4.271 & 2.772 & 4.1 \\
TV  & 1.125 & 0.623 & 0.9 \\
\hline
\end{tabular}
\caption{Comparison of 99th-percentile velocities across CFD, 4D Flow MRI, and echocardiography (Echo) at the left PA, right PA, and E-wave tricuspid valve inflow.}
\label{tab:echo}
\end{table}

Using localized velocity extraction at anatomically matched sites, we compared the 99th-percentile (p99) velocity from the CFD field and the 4D flow MRI-derived velocity field against the available echocardiographic peak velocity measurements at the left PA, right PA, and TV (Table \ref{tab:echo}). For the branch PAs, sphere-based sampling yielded CFD p99 values of 3.717 m/s in the LPA and 4.271 m/s in the RPA, compared with 4D flow MRI values of 2.297 m/s and 2.772 m/s, and echo peak velocities of 4.2 m/s and 4.1 m/s, respectively. Thus, the CFD results were in closer agreement with the echo in the branch PAs, whereas the 4D flow MRI velocities were systematically lower. At the TV, the CFD p99 velocity was 1.125 m/s, which was higher than the echo E-wave peak velocity of 0.9 m/s, whereas the 4D flow MRI p99 velocity was lower at 0.623 m/s.

\begin{figure}[h!]
    \centering
    \includegraphics[width=\linewidth]{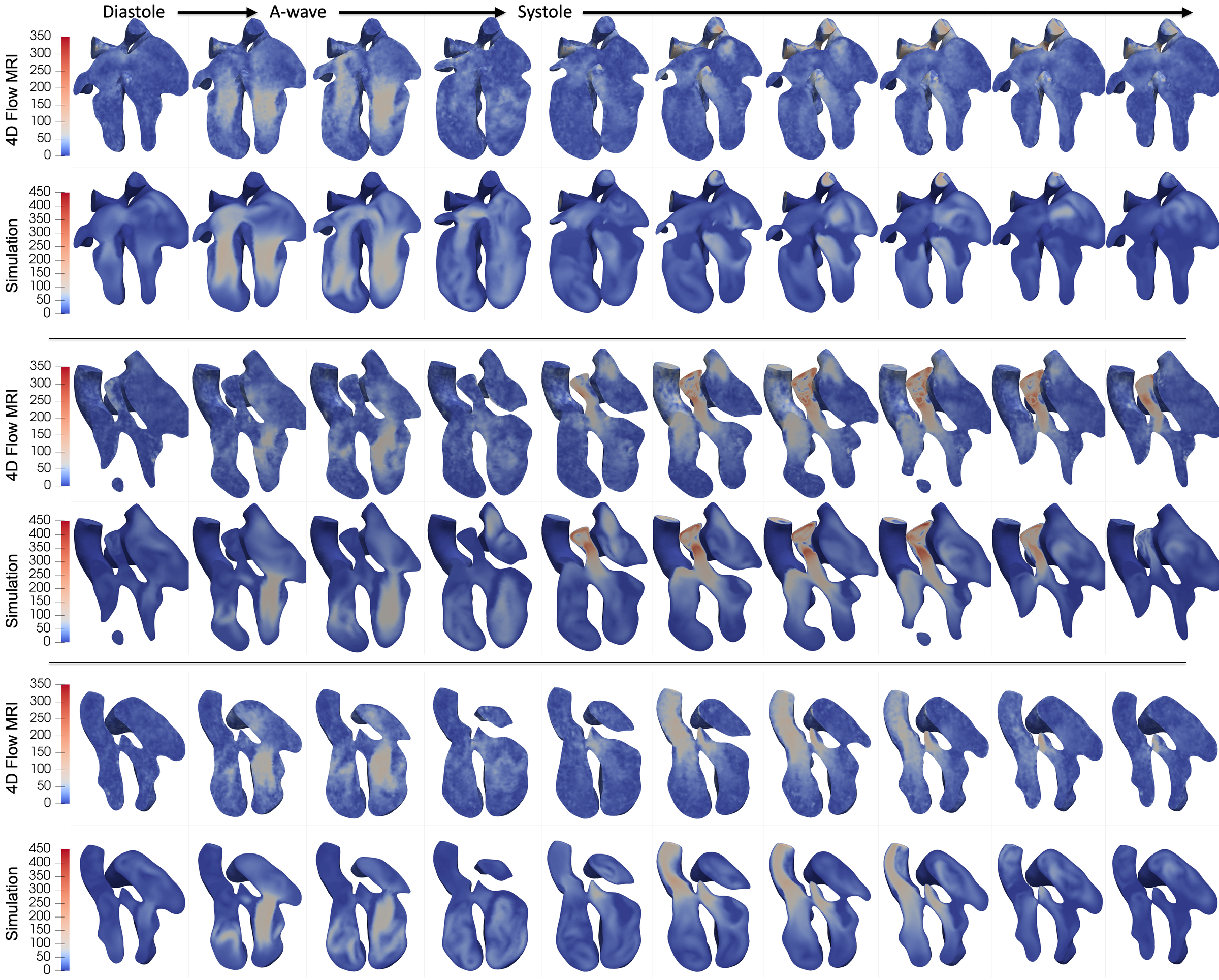}
    \caption{Comparison of simulated velocity fields with 4D-Flow MRI. The comparison is performed at three cross-sectional slices corresponding to the four-chamber view, pulmonary artery, and aorta at different phases of the cardiac cycle. Color scales are adjusted separately for the simulation and 4D-Flow MRI to better highlight flow features in each. }
    \label{fig:4dflow_slices}
\end{figure}

Figure \ref{fig:4dflow_slices} compares the simulated velocity fields with 4D-Flow MRI measurements at three representative cross-sectional planes: the four-chamber view, the pulmonary artery, and the aorta. Across all phases of the cardiac cycle, the RIS–ALE simulation captured the overall flow structures and temporal evolution observed in the MRI data. During early diastole, the inflow jets entering the ventricles through the atrioventricular valves exhibited similar orientations and orifice areas in both cases. During systolic ejection, the simulation reproduced the high-velocity outflow jet and its pattern in the pulmonary and aortic tracts, including the strong flow acceleration through the pulmonary artery consistent with the patient’s pulmonary stenosis, and the subsequent jet impingement at the pulmonary bifurcation, which generated localized flow separation and vortical recirculation. The simulation also captured the shunting through the VSD, showing flow from the morphologic right ventricle into the pulmonary artery during late systole, while the VSD flow shunting was less apparent during early contraction in both the simulation and 4D-Flow MRI.

While the simulation generally matched the flow directions and timing of velocity peaks, the 4D-Flow MRI data exhibited smoother velocity gradients and slightly lower peak magnitudes. Nevertheless, this may be due to the limited spatial and temporal resolution of 4D-Flow MRI, as well as measurement artifacts associated with the abnormally high inflow velocity into the PA. The simulation additionally revealed more detailed intraventricular vortical structures, flow through the VSD and high-speed flow patterns in the PA, features that were partially obscured in the MRI due to noise and partial volume effects. Overall, the comparison demonstrates strong qualitative agreement between the simulated and measured flow fields, confirming that the RIS–ALE framework accurately reproduces the patient-specific intracardiac flow patterns observed in vivo.

\begin{figure}[h!]
    \centering
    \includegraphics[width=\linewidth]{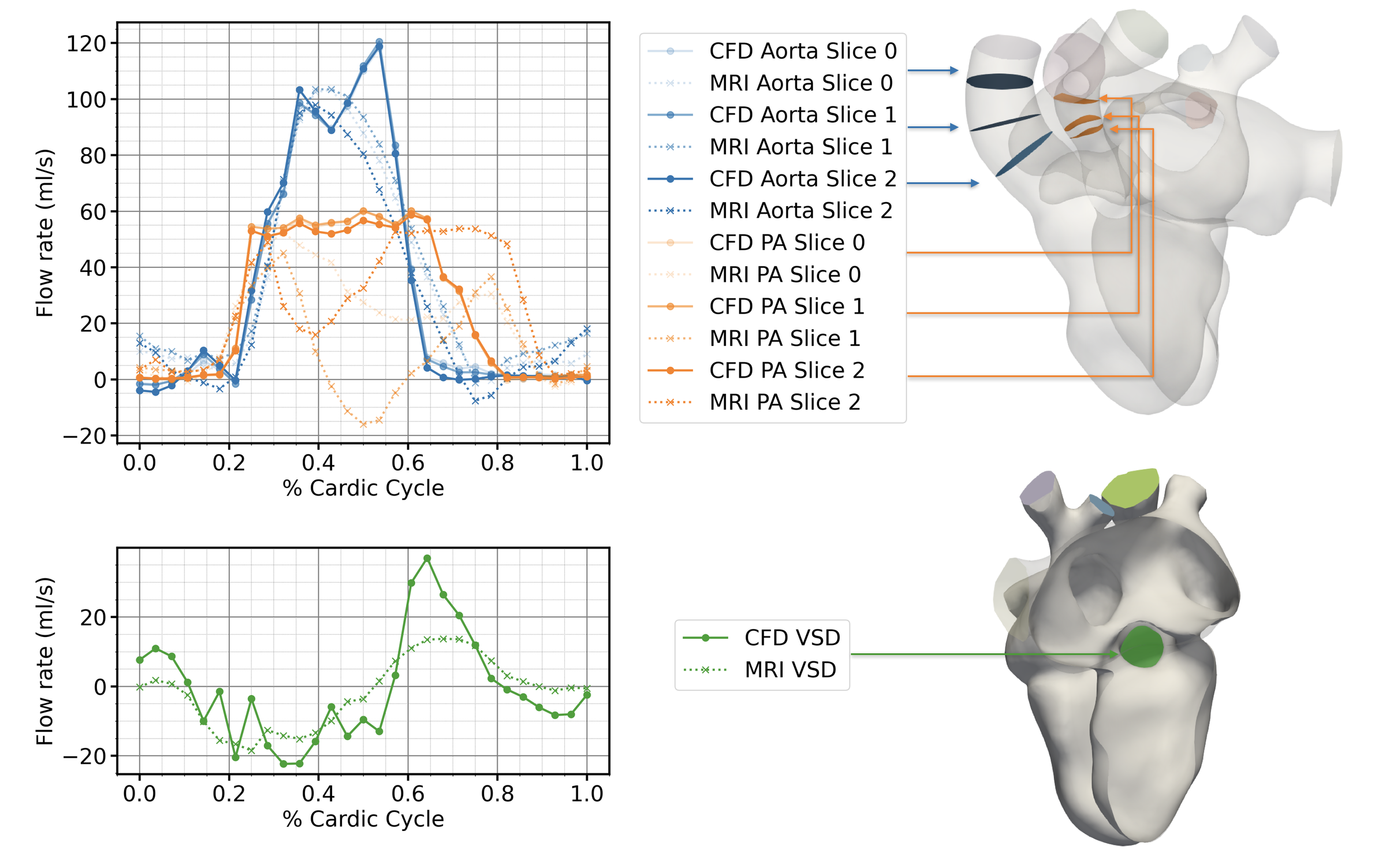}
    \caption{Comparison of simulated and MRI-derived flow rates over the cardiac cycle. Flow rates in the aorta and PA (top) and shunt flow across the VSD (bottom) are compared between CFD simulations and 4D-Flow MRI. Flow rates are evaluated over one cardiac cycle. The solid lines represent flow rates derived from the simulations, while the dashed lines represent flow rates derived from 4D-Flow MRI. Different colors indicate flow rates evaluated at PA, Aorta or VSD. The right panel shows the cross-sectional planes used to compute the aortic and PA flow rates, as well as the VSD surface where shunt flow is evaluated.}
    \label{fig:q_compare_cfd_mri}
\end{figure}

Figure~\ref{fig:q_compare_cfd_mri} presents the flow rates in the PA and the aorta, as well as the shunt flow across the VSD, comparing results from the CFD simulations with those obtained from 4D-Flow MRI. Flow rates were evaluated at discrete time points corresponding to the available 4D-Flow MRI frames. The PA and aortic flow rates were computed by integrating the velocity component normal to cross-sectional planes placed within the respective outflow tracts. To consider the effect of noise and any imaging artifacts in the 4D-Flow MRI data, flow measurements were repeated at three nearby cross-sectional locations along each outflow tract, and the resulting variability is reported by plotting all flow rate curves. As shown in Figure~\ref{fig:q_compare_cfd_mri}, the simulated aortic flow rate closely agrees with the MRI-derived flow rate in both magnitude and temporal trend. In contrast, the PA flow rate extracted from 4D-Flow MRI exhibits substantial variability across the three measurement locations, which, likewise, may be attributed to the small diameter of the PA and the presence of a high-velocity outflow jet that is difficult for MRI to capture. Despite this variability, the simulated PA flow rate during systole is in reasonable agreement with the maximum systolic PA flow observed in the MRI data. For flow shunting across the VSD, the simulations successfully capture left-to-right shunting from the morphological LV to the morphological RV during the first half of the cardiac cycle, followed by reverse shunting later in the cycle.


\section{Discussion}

We developed an image-based, four-chamber cardiac flow modeling framework that combines moving-domain finite element CFD with RIS valves to enable patient-specific interrogation of intracardiac hemodynamics in both normal and CHD hearts. Patient geometries and wall motion were reconstructed from time-resolved CTA (healthy adult) and 4D-Flow MRI magnitude images (CHD) using ML segmentation, followed by diffeomorphic mesh propagation via a neural ODE registration model. The resulting deforming fluid domain was simulated using an ALE Navier–Stokes formulation in svFSI and coupled to a closed-loop 0D lumped-parameter circulation to provide physiologic pressure boundary conditions. All four cardiac valves were incorporated using RIS, with valve opening and closure triggered by pressure-gradient and backflow criteria, allowing efficient representation of valve function without using fully coupled FSI. The simulations reproduced physiologic chamber pressure–volume behavior and flow patterns in the healthy case. In the CHD case, the simulated pressure matched catheterization-derived pressures within approximately 10\% for most chambers and great vessels. The simulation also captured the abnormally elevated systolic pressure in the functional RV, which is nearly equal to the systolic pressure in the functional LV, as well as the abnormally large pulmonary valve pressure gradient. Compared with 4D-Flow MRI, the CHD simulations showed strong qualitative agreement in the timing and orientation of diastolic inflow jets, systolic ejection into the aorta and pulmonary artery, and flow features related to narrowed PA and VSD, while providing higher-resolution visualization of flow structures that are partially obscured by MRI noise and limited spatiotemporal resolution. These results demonstrate that our ALE-RIS-based simulation framework enables computationally efficient yet physiologically realistic simulations that can recover detailed intracardiac flow patterns within patient anatomy.

We first highlight the advantages of our framework. Our approach enables detailed characterization and quantitative analysis of intracardiac flow at high spatial and temporal resolution. As shown in Figures~\ref{fig:4dflow_slices} and~\ref{fig:q_compare_cfd_mri}, the 4D-Flow MRI measurements were very noisy and yielded flow-rate estimates that varied across nearby analysis planes. Indeed, 4D-Flow MRI is limited by relatively coarse spatial (1.5-2.5mm) and temporal resolution (>20ms), as well as substantial noise and motion artifacts ~\cite{bissell_4d_2023, demirkiran_clinical_2022, zhuang_role_2021}. These limitations introduce substantial errors in derived hemodynamic metrics, particularly near walls and in regions of complex flow \cite{ha_assessment_2016, schmidt_impact_2021, zhuang_role_2021}. These challenges are further exacerbated in our case by the small size of the great vessels and the rapid heart rate of the infant, which lead to significant variability in the measured flow fields.  By contrast, our CFD simulations provide spatially and temporally consistent velocity and pressure fields, and detailed intracardiac flow structures that are difficult to recover from 4D-Flow MRI alone.

Nevertheless, since intracardiac flow is strongly governed by patient-specific anatomy and cardiac motion, the fidelity of the simulated flow critically depends on accurate reconstruction of the moving cardiac geometry from time-series image data. This requirement is addressed in our framework through automated and semi-automated ML-based segmentation combined with diffeomorphic mesh propagation. While many prior image-derived CFD studies have been limited to a single chamber~\cite{kong_automating_2020, loke_computational_2022, schenkel_mri-based_2009}, such as the left ventricle, or to the left heart alone~\cite{chnafa_image-based_2014,vedula_effect_2016}, we reconstruct the entire four-chamber domain together with the major connected vessels, including the aorta, pulmonary artery, pulmonary veins, and venae cavae, thereby enabling comprehensive four-chamber hemodynamic analysis. Notably, although constructing a temporally resolved full-heart mesh is traditionally labor-intensive, the ML-based segmentation and neural ODE deformation model substantially reduces this burden: after generating a single high-quality baseline mesh and defining faces for assigning boundary conditions, the remaining time frames are obtained by automatically propagating the mesh through the cardiac cycle. The diffeomorphic NODE formulation enforces smooth, non-intersecting trajectories of mesh points, yielding temporally consistent deformations that preserve mesh topology while capturing large cardiac motion. As demonstrated in Figures \ref{fig:healthy_seg} and \ref{fig:vsd_seg}, our deformed mesh closely matched tissue boundaries on the patient image data throughout the cardiac cycle. 

A major advantage of the ALE–RIS framework is that it substantially reduces computational cost while still accounting for physiologically realistic valve dynamics. In contrast to approaches that solve a monolithic FSI problem, coupling blood flow and deforming valve leaflets, our framework decouples valve mechanics from the fluid solver. Valve opening, closure, and leaflet contact are computed separately using shell-based contact simulations, and their hemodynamic effects are then incorporated into the ALE Navier–Stokes equations through the RIS formulation. This strategy avoids the high computational cost associated with simultaneously solving fluid and structural equations for the valves, while still capturing the dominant influence of valve dynamics on intracardiac flow. For the CHD case, four cardiac cycles were completed in approximately 18 hours using three computing nodes with 16 cores per node, and for the healthy case, two cardiac cycles were completed in approximately 19 hours on Intel Xeon Gold 5118 (Skylake) CPUs. These runtimes are substantially lower than those typically reported for whole-heart FSI simulations with explicit valve mechanics \cite{feng_whole-heart_2024}. 

Next, we discuss the simulated cardiac flow results for the CHD patient. To match the catheterization-derived pressures, it was necessary to substantially increase the systemic arterial resistances, with the systemic arterial resistance raised from 0.677 to 1.578~mmHg·s·mL$^{-1}$ (approximately 11 to 26 Wood units, WU). The resulting systemic arterial resistance lies within the reported range of 15–30 WU for a 15-month-old child whose systemic vascular resistance has approached near-adult levels~\cite{souza_cardiovascular_2011}. During parameter tuning, we found that this increase in resistance was required to reproduce the observed systolic flow partitioning between the aorta and the PA. In particular, when the systemic arterial resistance was set too low, an excessive fraction of flow was diverted into the aorta, because the PA is narrower than the aorta in this patient, and the presence of a VSD allows flow from both ventricles to be directed into either the aorta or the PA. In addition, we increased the pulmonary arterial resistance from 0.032 to 0.136~mmHg·s·mL$^{-1}$ (0.53 to 2.27 WU) when transitioning from the healthy to the CHD patient. This value is consistent with the pulmonary vascular resistance measured by cardiac catheterization (2.3 WU) in this patient. With this resistance, the LPN reproduced physiologic periodic pulmonary arterial flow as well as pulmonary arterial and venous pressures. Furthermore, we observed a decrease in both systemic and pulmonary arterial capacitance when transitioning from the healthy adult to the CHD patient, although blood vessels in a 15-month-old child are intrinsically more compliant than those of adults \cite{gardner_association_2010}. In the LPN framework, however, arterial capacitance represents the effective volume storage capacity of the vasculature rather than vessel wall stiffness alone. Accordingly, the reduced capacitance reflects the substantially smaller blood volume and stroke volume that the CHD pediatric patient's vascular system accommodates. Lastly, we found it necessary to significantly increase the capacitance values and decrease the inductance values in the arterial and pulmonary venous systems. These adjustments were introduced to smooth high-frequency flow rate fluctuations in the pulmonary veins and vena cava within the LPN. Because the flow rates in these vessels are prescribed from the 3D simulations, we believe these fluctuations arise from temporal inconsistencies in the prescribed motion, likely due to a combination of imaging noise and segmentation and registration errors in those small vessels.

When comparing the CFD results with the 4D-Flow MRI measurements, the simulated flow patterns qualitatively resembled those observed in the MRI data but generally exhibited higher velocity magnitudes, as shown in Figures~\ref{fig:4dflow_volume} and~\ref{fig:4dflow_slices}. Because the velocity magnitude is primarily governed by the flow rate through the chambers driven by cardiac motion, which was directly derived from the magnitude images of the 4D-Flow MRI, the discrepancy is unlikely to arise from errors in the CFD simulation itself. Instead, the apparent underestimation in the MRI measurements likely reflects limitations of 4D-Flow MRI, including its relatively coarse spatial and temporal resolution, which can average high-velocity jets over multiple voxels and cardiac phases. In addition, the velocity-encoding setting (VENC) used during MRI acquisition may be lower than the actual peak velocities in the pulmonary outflow tract, which are abnormally high for this CHD patient. This mismatch can lead to velocity aliasing and further contribute to the reduced velocity magnitudes observed in the measured data.

In healthy hearts, late-diastolic filling in the LV is commonly organized by the formation of a dominant intraventricular vortex. This vortex emerges from the interaction between mitral inflow, ventricular geometry, and coordinated wall motion, and has been consistently reported in both imaging and computational studies \cite{pedrizzetti_vortexearly_2014, kilner_asymmetric_2000, pedrizzetti_nature_2005,seo_effect_2013,cimino_vivo_2012}. Prior work has suggested that such coherent vortex organization supports efficient redirection of inflow toward the outflow tract and is associated with reduced viscous dissipation, thereby facilitating an energetically favorable transition from diastole to systole \cite{pedrizzetti_nature_2005, seo_effect_2013, cimino_vivo_2012, davies_abnormal_1992}. In the present ccTGA case, this canonical vortex organization in the systemic ventricle appears to be disrupted as a consequence of ventricular–arterial discordance. Because systemic ejection is performed by the morphological RV, the intraventricular flow more closely resembles RV-like vortex topology, with multiple vortical structures rather than a single coherent vortex. In addition, in the morphological LV connected to the PA, late-diastolic circulation is reversed relative to the healthy LV, with inflow initially directed away from the outlet before being redirected toward ejection. From an energetic perspective, these altered vortex patterns are accompanied by a higher ratio of volume-normalized viscous dissipation to kinetic energy compared with the healthy subject, suggesting less efficient preservation of organized flow (Figure \ref{fig:energy}; see Appendix for details of the computation). Similar associations between disturbed ventricular vortex formation and increased viscous energy loss have been reported in patients with corrected atrioventricular septal defects \cite{elbaz_impact_2015}. The more pronounced increase in this ratio during systole in the CHD patient as shown in Figure \ref{fig:energy} may reflect the combined effects of abnormal ventricular morphology, altered alignment with the great vessels, a narrowed pulmonary outlet, and the presence of a large VSD. Elevated systolic energy losses have also been observed in Fontan patients compared with controls, and particularly in Fontan cohorts with discordant inflow–outflow configurations \cite{kamphuis_disproportionate_2019}. Together, these observations suggest that inefficient flow organization and elevated energy loss may occur across various congenital conditions, and that CFD enables three-dimensional, time-resolved characterization of flow energetics that is difficult to obtain from imaging alone.

\begin{figure}[h!]
    \centering
    \begin{subfigure}[b]{\textwidth}
        \centering
        \includegraphics[width=\textwidth]{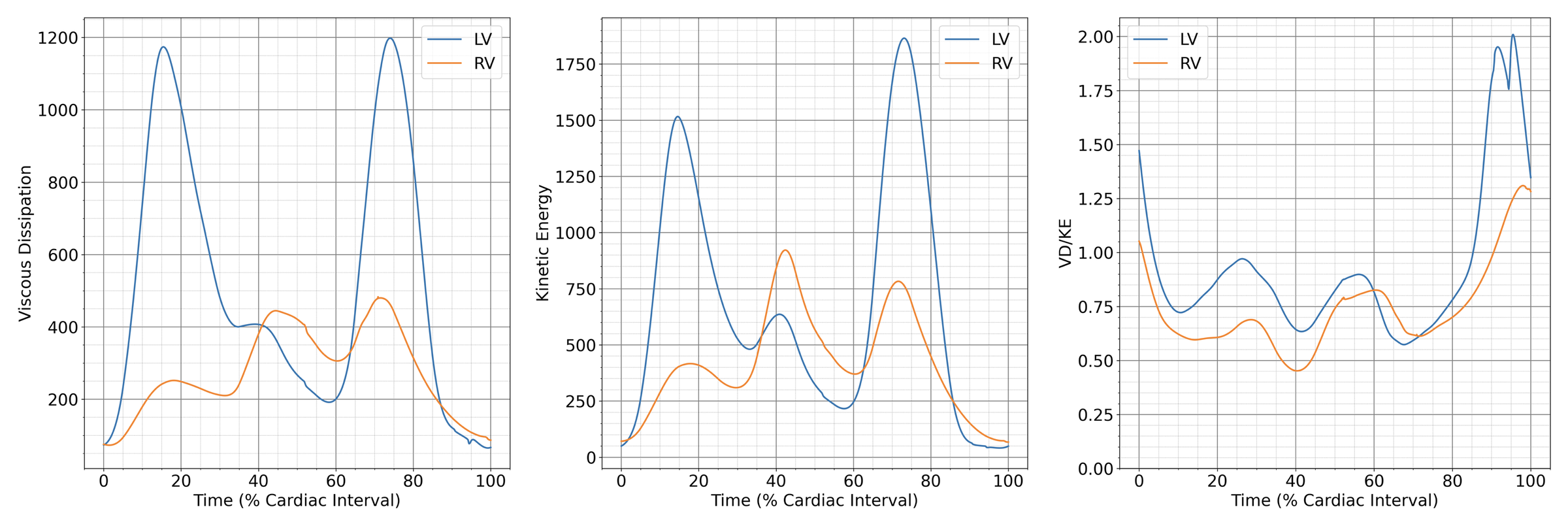}
        \caption{Healthy subject}
        \label{fig:healthy_energy}
    \end{subfigure}
    \hfill 
    \begin{subfigure}[b]{\textwidth}
        \centering \includegraphics[width=\textwidth]{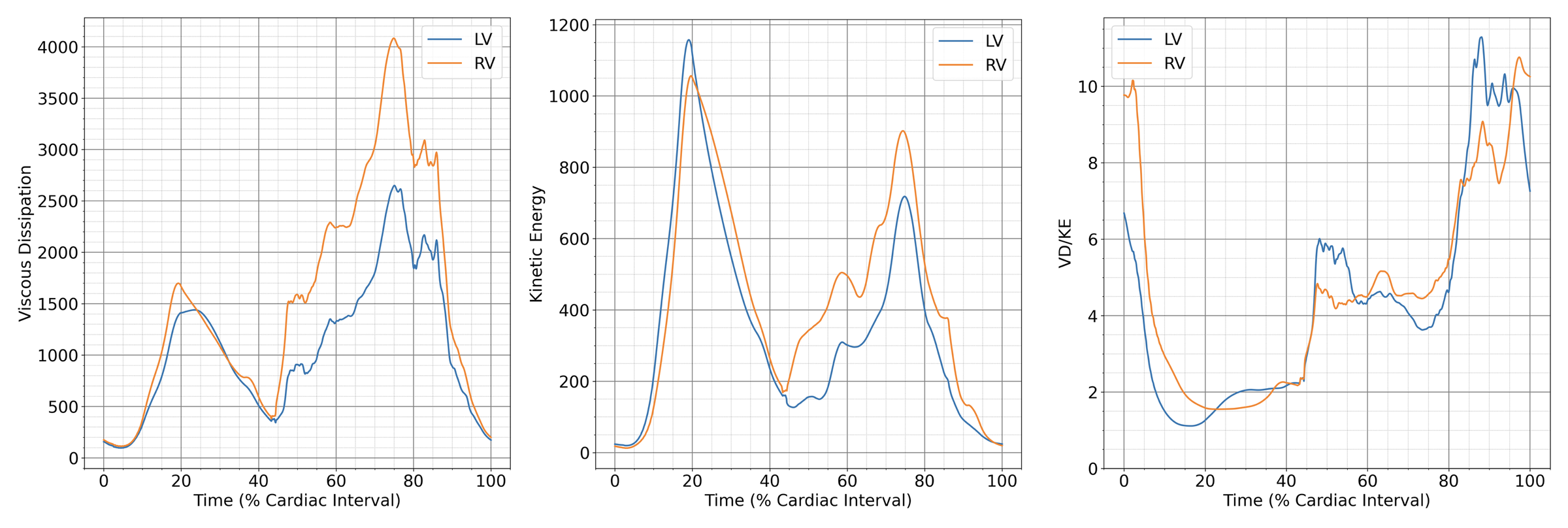}
        \caption{CHD patient}
        \label{fig:vsd_energy}
    \end{subfigure}
    \caption{Comparison of volume-normalized viscous dissipation ($\mathrm{dyne/(cm^2 \cdot s)}$), kinetic energy ($\mathrm{dyne/cm^2}$), and the ratio of viscous dissipation to kinetic energy ($\mathrm{s^{-1}}$) between the healthy subject and the CHD patient. The cardiac cycle is plotted starting at the onset of diastole. For the CHD patient, LV and RV refer to the functional left and right ventricles. }
\label{fig:energy}
\end{figure}


Our ALE-RIS framework for whole heart cardiac flow simulations has the following limitations. The accuracy of the simulated pressure transients and valve timing is limited by the temporal resolution of the image-derived motion. In both cases, wall motion was prescribed at discrete image frames and then interpolated in time, which can smooth or distort rapid events near end-diastole and end-systole. This limitation is most evident in the unrealistically short isovolumetric contraction and relaxation phases, which are on the order of $\sim 10$ ms in our simulations compared with physiological values of $70-100$ ms \cite{harrison_relation_1964,benchimol_study_1967}. Because the wall motion is not constrained to be strictly volume-preserving during these phases, small interpolated volume changes can occur when both inlet and outlet valves are effectively closed, leading to spurious pressure spikes, kinks in atrial and ventricular pressure traces, and oscillations in the PV loops. These artifacts are also related to the observed pressure overshoot during isovolumetric contraction and the mild undershoot during isovolumetric relaxation. Several strategies could reduce these errors in future work. First, the interpolation scheme could be replaced by a temporally regularized motion model that penalizes non-physiological accelerations of the meshes over time. Second, additional constraints could be imposed during isovolumetric phases, for example, enforcing near-constant ventricular volume (or enforcing a target volume curve) when both valves are closed, to prevent spurious compression/expansion that drives artificial pressure transients. Lastly, although computationally expensive, fully coupled multiphysics FSI simulations that integrate electrophysiology, cardiac mechanics, and fluid dynamics could be used to generate more physiological heart motion. By driving contraction through electrophysiology and incorporating physiological fiber architecture, such models can better capture important nuances of cardiac kinematics than prescribed, image-interpolated wall motion.

In addition, the reconstructed cardiac motion in this study is fully derived from image-based segmentations and is therefore limited by the information encoded in image intensity. As a result, ventricular torsional motion, which arises from the coordinated contraction of helically organized myocardial fibers with transmural variation in fiber orientation, is not explicitly captured, since this motion cannot be resolved from CT or 4D-Flow MRI alone. While this represents a limitation of the current approach, prior studies suggest that the omission of torsion is unlikely to substantially affect the intracardiac flow patterns examined here. Vasudevan \etal reported that ventricular torsion resulted in minimal changes to flow structures and contributed less than 2\% to energy loss, wall shear stress, and ejection momentum in their simulations \cite{vasudevan_torsional_2019}. Similarly, Miyauchi \etal found that torsional motion had a negligible effect on average relative residence time, a metric of blood stagnation, in LV models that did not include trabeculae or papillary muscles \cite{miyauchi_numerical_2023}.

Lastly, we note that patient-specific valve leaflet geometries were not directly reconstructed from imaging; instead, template geometries were used for all four valves and morphed to fit within each patient’s anatomy. Despite this simplification, the simulated flow fields in the CHD case showed good qualitative agreement with 4D-Flow MRI, suggesting that the dominant intracardiac flow features are primarily governed by chamber geometry, wall motion, and valve timing in patients without significant valvular disease, and can therefore be captured reasonably well even with simplified valve representations. However, in cases involving abnormal atrioventricular valves, valve repair or replacement, or valve regurgitation, intraventricular flow patterns may be significantly altered \cite{pugliese_flow_2023, lantz_impact_2021, faludi_left_2010}, and patient-specific valve anatomies, along with fully coupled FSI, would be required for more accurate modeling.







\section{Conclusion}
We presented an image-based, patient-specific framework for simulating whole-heart, four-chamber intracardiac hemodynamics that balances physiologic fidelity with computational efficiency in both simulation and model construction. The approach combines (i) ML-based segmentation and diffeomorphic neural-ODE mesh propagation to efficiently reconstruct temporally consistent moving cardiac anatomies from time-resolved imaging, (ii) ALE-based deforming domain CFD simulation of intraventricular flow, (iii) RIS to incorporate all four valves with pressure- and backflow-triggered opening and closure using realistic valve geometries, and (iv) a closed-loop 0D lumped-parameter network to represent systemic and pulmonary circulations and provide physiologic boundary conditions. The framework was demonstrated in both a healthy subject and a pediatric patient with complex CHD. In the healthy case, the simulations reproduced realistic chamber pressures and physiologic flow patterns. In the CHD case, simulated pressures agreed well with cardiac catheterization measurements, and the simulated flow patterns were qualitatively consistent with 4D-Flow MRI while providing higher spatial and temporal resolution flow fields. The model also captured key hemodynamic abnormalities, including elevated ventricular pressures, the elevated pressure gradient across the pulmonary valve, and abnormal intraventricular flow patterns in the CHD patient. Overall, these results support the use of the proposed ALE–RIS framework as a computationally efficient approach to extract detailed, patient-specific intracardiac flow information beyond what is accessible from current clinical imaging alone.

\section*{Acknowledgments}

This work was supported in part by NSF 1663671, NIH R01EB029362, NIH R01LM013120, NIH R38HL143615, NIH R01HL171515, and NIH R01HL173845. AB acknowledges the National Institutes of Health (grant numbers 5R01HL159970 and 5R01HL129727). LS acknowledges NSF 2501698. We would like to acknowledge Dr. Vijay Vedula for his helpful suggestions, and Dr. Fannie Gerosa for her help in RIS implementation.

\section*{Appendix}

At each time step $t$, the kinetic energy density and viscous dissipation density were computed from the simulated velocity field. The kinetic energy density was defined as
\begin{equation}
e_{\mathrm{KE}}(\mathbf{x},t) = \frac{1}{2}\rho \|\mathbf{v}(\mathbf{x},t)\|^2 ,
\end{equation}
where $\rho$ is the fluid density and $\mathbf{v}$ is the velocity vector.

Viscous dissipation density was computed from the strain-rate tensor
\begin{equation}
\mathbf{S}(\mathbf{x},t) = \frac{1}{2}\left(\nabla \mathbf{v}(\mathbf{x},t) + \nabla \mathbf{v}(\mathbf{x},t)^T\right),
\end{equation}
as
\begin{equation}
\phi(\mathbf{x},t) = 2\mu\,\mathbf{S}:\mathbf{S},
\end{equation}
where $\mu$ is the dynamic viscosity.

For each cardiac chamber $\Omega(t)$, the chamber-averaged kinetic energy density and viscous dissipation density were obtained by integrating the local quantities over the chamber volume and normalizing by the instantaneous chamber volume:
\begin{align}
\overline{e}_{\mathrm{KE}}(t) &= \frac{1}{|\Omega(t)|}\int_{\Omega(t)} \frac{1}{2}\rho \|\mathbf{v}\|^2 \, dV, \\
\overline{\phi}(t) &= \frac{1}{|\Omega(t)|}\int_{\Omega(t)} 2\mu\,\mathbf{S}:\mathbf{S} \, dV.
\end{align}

\bibliographystyle{unsrt}  
\bibliography{references}

\end{document}